\documentclass[11pt, oneside, reqno]{amsart}

\usepackage{color}
\usepackage[english]{babel}
\usepackage[utf8]{inputenc}
\usepackage{amsmath, amssymb}
\usepackage{url}
\usepackage{graphicx,colortbl}
\usepackage{bm}
\usepackage[margin=1in]{geometry}
\usepackage[colorlinks=true,citecolor=magenta,linkcolor=blue]{hyperref}
\definecolor{lightgray}{gray}{0.7}

\def\be#1\ee{\begin{equation}#1\end{equation}}

\newcommand{\bq}{\begin{equation}}
\newcommand{\eq}{\end{equation}}
\def\bqa{\begin{eqnarray}}
\def\eqa{\end{eqnarray}}

\newcommand{\bd}{\begin{displaymath}}
\newcommand{\ed}{\end{displaymath}}
\newcommand{\ba}{\begin{eqnarray}}
\newcommand{\ea}{\end{eqnarray}}

\newcommand{\eps}{\varepsilon}
\newcommand{\R}{\mathbb{R}}
\newcommand{\dd}{\mathrm{d}}
\newcommand{\sta}{\mathsf{static}}
\newcommand{\dyn}{\mathsf{dynamic}}

\usepackage{verbatim}

\begin{document}

\title[Three-layer model for flow of sedimenting suspensions]{A three-layer model for the flow of\\particulate suspensions driven by sedimentation}

\author[A. Bondesan]{Andrea Bondesan}
\address{Andrea Bondesan \hfill\break
	Department of Mathematical, Physical and Computer Sciences, University of Parma \hfill\break
	I-43124 Parma, Italy}
\email{andrea.bondesan@unipr.it}

\author[L. Girolami]{Laurence Girolami}
\address{Laurence Girolami \hfill\break
	RECOVER, INRAE, Aix--Marseille Université \hfill\break 
	F-13182 Aix-en-Provence, France \hfill\break
	Laboratoire GéHCO - EA 6293, Université de Tours \hfill\break
	F-37200 Tours, France}
\email{laurence.girolami@univ-tours.fr}

\author[F. James]{François James}
\address{François James \hfill\break
	Institut Denis Poisson - UMR 7013, Université d’Orléans \hfill\break
	F-45067 Orléans, France}
\email{francois.james@math.cnrs.fr}

\author[L. Rousseau]{Loïc Rousseau}
\address{Loïc Rousseau \hfill\break
	Laboratoire GéHCO - EA 6293, Université de Tours \hfill\break
	F-37200 Tours, France}
\email{loic.rousseau@univ-tours.fr}


\begin{abstract}
We introduce a system of shallow water-type equations to model laboratory experiments of particle-laden flows. We explore homogeneous liquid-solid suspensions of fine, non-cohesive, monodisperse glass beads which propagate as an equivalent fluid that progressively sediments and forms a growing deposit at the bottom of a smooth channel, and simultaneously creates a thin layer of pure liquid at the surface. The novelty of this model is twofold. First, we fully characterize the first-order behavior of these flows (mean velocity, runout distances and deposits geometry) through the sole sedimentation process of the grains. The model remains very simple and turns out to be effective despite the complex nature of interactions involved in these phenomena. Secondly, the sedimentation dynamics of the grains is observed to not be strongly affected by the flow, remaining comparable to that measured in static suspensions. The mathematical model is validated by comparing the experimental kinematics and deposit profiles with the simulations. The results highlight that this simplified model is able to describe the general features of these flows as well as their deposit morphology, provided that the settling rate is properly adjusted from a threshold Reynolds number, i.e. when the flow becomes sufficiently agitated to disturb and delay the deposition processes. 
\end{abstract}

\maketitle

\vspace*{0.5cm}
\hspace*{2.1cm} Multi-layer shallow water model; Hydrostatic upwind finite volume scheme.

\vspace*{0.5cm}


\section{Introduction} \label{sec:introduction}

Lahars are geophysical mass flows, made with hot volcanic ash and tephra or fluvial sediments suspended into water, that travel down valleys at high velocity under the influence of gravity. During propagation, their general behavior undergoes important modifications in particle concentration \cite{ValIve}, thus evolving from slightly concentrated flows (with a solid concentration lower than $20\ \textrm{vol}\%$), to hyperconcentrated flows (with a solid concentration ranged from $20\ \mathrm{vol}\%$ to $50\ \mathrm{vol}\%$) or debris flows (with a solid concentration that exceeds $50\ \mathrm{vol}\%$) \cite{PieSco}. Such large-scale natural phenomena are highly hazardous and usually lead to important damages \cite{Gud}. With the aim of providing reliable tools to help local authorities to establish proper risk assessments and take relevant preventing decisions, it appears fundamental to understand the physical processes that govern the flow dynamics and to develop predictive models of their runout time, distance and deposit geometry. Dam-break experiments involving particulate suspensions consist in the sudden release of a fluidized granular column, of known concentration, down a horizontal plane and provide a simplified way to describe such complex geophysical events. The study of canonical configurations of small-scale collapses thus became largely explored in the last decades, leading to the development of predictive laws guided by two complementary approaches. 

\smallskip
The first one is based on a dimensional analysis which allows to identify the main physical parameters governing the flow dynamics, as a mean of deducing empirical laws that involve non-dimensional groups able to describe the flow propagation (runout time and length, flow velocity) and their deposits. In recent years, small-scale experiments of granular slumping performed under controlled conditions in both axisymmetric and rectangular configurations have highlighted the description of the flow dynamics and deposit geometry in terms of the free-fall collapse, described by the gravity time and speed, as well as the initial geometry of the column, described by its aspect ratio \cite{LajManVil, LHSH}. Similar investigations of moderately expanded ($2.5$-$4.5\%$) gas-particles systems have exhibited the water-like behavior of these mixtures dominated by an excess of fluid pressure, and have also shown the dependency on the initial geometry of the slightly fluidized column \cite{RMNT, Roc}. Additional experiments of highly expanded (up to $50\%$), fully fluidized, dam-break suspension flows have been conducted, highlighting the effect of the solid concentration on their dynamics dominated by the sedimentation processes that control the emplacement of such experimental and natural flows \cite{GirDruRoc, GirRocDruCor}. In particular, the sedimentation velocity of particles was observed to not be affected by the flow, thus remaining comparable to that inferred from static suspensions which results from simple vertical collapse tests \cite{GirDruRoc, GirRis2}. Afterwards, dam-break avalanches of wet granular materials, involving both unsaturated and saturated mixtures, were developed and underlined the dependency on the particle concentration, as well as on the Bond and Stokes numbers, to describe the column collapse and their dynamics \cite{ASGTC, BouLacBon}.

\smallskip
The second approach exploits these dimensional analyses, primarily established for the development of experimental scaling laws, to define the proper physical assumptions underlying the derivation of partial differential equations of mass and momentum conservation, which can be used to model and simulate the flows \cite{LajMonHom, DHLMS, HutSveRic, Ive, ManMajFag}. Since the dynamics of geophysical mixtures may involve several complex mechanics such as dilatancy effects, inter-particle and fluid-grain friction, sedimentation, resuspension, particle segregation, it is important to identify the main processes of interest, depending on the phenomenon and on the scientific issues under consideration. Many scientific efforts devoted to the mathematical description of such phenomena have been developed in recent years, based on the leading idea that the flow evolves as a thin homogeneous layer in which horizontal motions are much greater than the vertical ones. The resulting models are thus obtained by depth-integrating Navier--Stokes/Euler equations under a thin-layer approximation. In particular, we can distinguish single-phase models \cite{Ost, Ive, BGGNS, MedBatHur, Fer_etal, DoyHogMad, ShiKoySuz} from quasi single-phase ones \cite{IveDen, PudWanHut, IveGeo}, where two equations are required for the mass conservation of both the mixture and the solid phase, while the relative motion and interactions between the fluid and the solid components are neglected; or with two-phase models \cite{PitLe, PelBouMan, PaiPou, Pud, KowMcE, BFMN1, BFMN2, LCHPL} where a system of four coupled equations for mass and momentum conservation of each phase are required to describe the evolution of the multi-phase flow. A distinct mention should also be made for the multi-component models \cite{MerTamRoc, BauKam} based on a very general framework that considers the full Navier--Stokes equations for both the fluid and the solid phases, avoiding any depth-averaged approximation.

\smallskip
In this work, we initially developed a series of small-scale laboratory experiments to study the propagation of homogeneous suspensions made with fine, non-cohesive, monodisperse glass beads and water, down a horizontal channel. The suspensions, whose solid volume fraction range from $0.41$ to $0.52$, are obtained by fluidizing a granular material in a reservoir. Previous experimental investigations highlighted that the flow dynamics is primarily controlled by sedimentation (initiated after the gravitational collapse down the rectangular flume) until all particles have deposited and the flow is at rest, which provides a natural criterion for the stopping phase \cite{GirRis2}. The sedimentation velocity appears to not be affected by the propagation, in the sense that it remains mostly constant with time and space and is well approximated by the settling speed of particles measured from the static regime. Starting from this analysis, we then derive simplified mathematical equations to model the experiments. We propose a multi-layer approach to describe the evolution of a suspension that behaves like an idealized equivalent fluid, forming a deposit that thickens with time and a superficial layer of water that is originated by the loss of mass due to sedimentation. In particular, the resulting system of shallow water equations is obtained by depth-integrating a coupled incompressible Euler--Navier--Stokes formulation that carefully accounts for the three evolving physical interfaces between the different layers. Similar multi-layer shallow water equations have been used in various geophysical contexts. We recall for example the application of two/three-layer models to describe sediment transport in rivers \cite{Tak}, the motion of stratified flows \cite{AbgKar, CFGP, CanFagLan, SCMT} and the dynamics of geophysical fluids like tsunamis \cite{Ost}, avalanches and debris flows \cite{LHKT}, submarine landslides \cite{Fer_etal} and pyroclastic currents \cite{DoyHogMad, ShiKoySuz}. Our model also accounts for the initial column collapse of the suspensions (where the shallow water approximation is not appropriate) by adapting a method developed for the slumping phase of dam-break granular flows \cite{LarStaHin, DHLMS}, interpreted as the free-fall of mass into a moving apparent thin layer.

\smallskip
The numerical treatment of this type of systems is challenging \cite{Par, AbgKar, BouZei, CFGP, CFNPP, BerFouMor}, since they involve non-conservative product terms and the eigenstructure of the full coupled equations cannot be determined explicitly. Moreover, the simple application of standard solvers accounting for the separate evolution of each layer usually leads to numerical instabilities. We chose to follow the strategy proposed by Berton et al. \cite{BerFouMor}, which is simple and efficient altogether, to design a well-balanced and positivity-preserving scheme. Specifically, we apply a suitable change of variables that allows to switch from the real layers heights to an apparent-depths formulation, transforming the system of equations into a new one which is appropriately decoupled (rather, the coupling is implicit in the new variables) and standard finite volume techniques can be applied to discretize the numerical fluxes in each layer separately and simulate the equations avoiding instabilities.

\smallskip
In order to predict the particle sedimentation rate from the mixture during propagation, we use a general law valid for suspensions of various concentrations (from the dilute regime to the denser ones) involving all types of materials and fluids \cite{Gir_etal}, when the fluid inertia remains negligible (i.e. in the Stokes flow regime). This law includes the theoretical formulation of a single particle falling into a homogeneous fluid of equivalent properties (in terms of mixture density and viscosity). Our approach is thus limited to homogeneous cases that can be described by a one-phase model and does not include processes like particles segregation, turbulent stresses or pore pressure effects \cite{Pud,KowMcE,LCHPL,BauKam}, which require a more general two-phase flow description, well suited to such heterogeneous suspensions, which is however beyond the scope of the present paper.

\smallskip
In addition to its simplicity, the proposed model is able to reproduce the first-order features of analog dam-break flow experiments, which can be viewed as small-scale hyper-concentrated flows or lahars. In this framework, the model provides an original point of view including a depositional description that controls the general behavior of such suspension flows during their final stage of transport, when dominated by sedimentation.

\smallskip
The work is structured as follows. We first present the experimental device and the laboratory experiments in Section 2. Based on these observations, we formulate the main physical assumptions in Section 3 and we introduce the mathematical model, as well as the numerical strategy for the simulations. Section 4 is devoted to the comparison between numerical results and experiments. We conclude with some final remarks in Section 5.

\begin{figure*}
\centering
\includegraphics[width=0.95\linewidth]{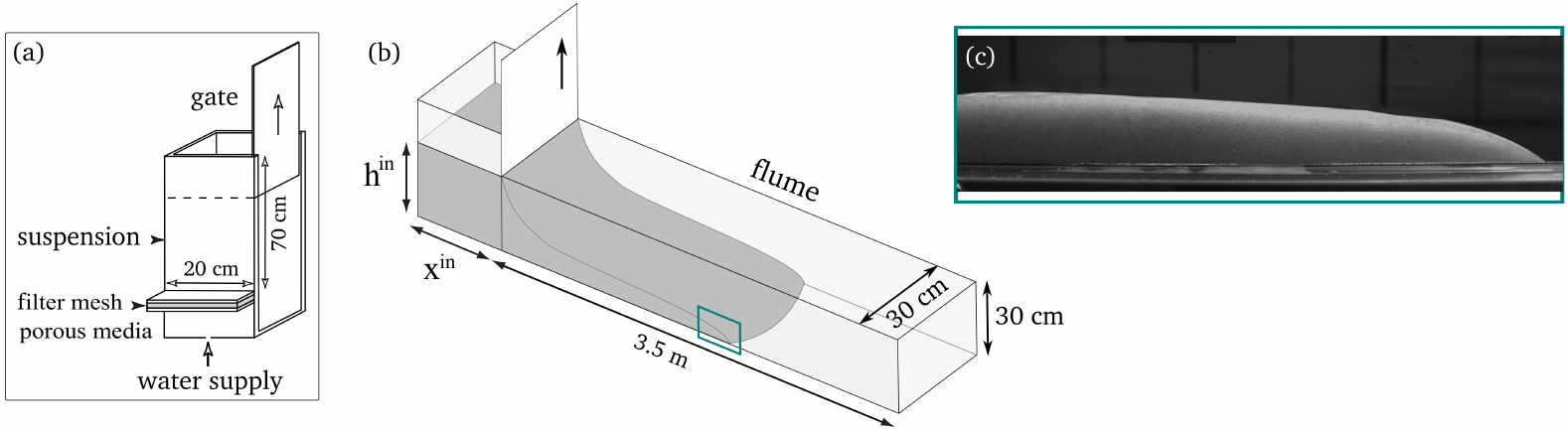}
\caption{\textbf{Experimental device}: (a) Lock-exchange reservoir where suspensions are generated using a fluidization technique. (b) Dam-break flume where suspension flows are generated and explored through the transparent side walls. (c) Picture of a suspension flow made with glass beads and water.}
\label{fig:experimental device}
\end{figure*}

\section{Description of the experiments} \label{sec:experiments}

\subsection{Experimental procedure}

\begin{figure}[h]
\centering
\includegraphics[width=0.5\linewidth]{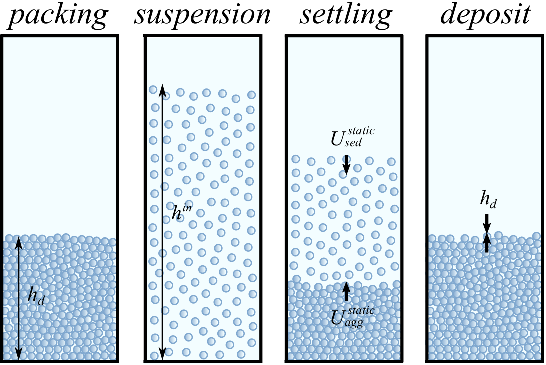}
\caption{\textbf{Sedimenting suspension}: Deposition processes from a homogeneous suspension \cite{Gir_etal}.}
\label{fig:fluidization}
\end{figure}

The experiments consist in generating homogeneous particulate suspensions in aid of fluidization techniques and to explore their propagation when flowing down a horizontal, impermeable channel (Figure \ref{fig:experimental device}). In doing so, solid particles of glass beads are first placed into a rectangular reservoir of $10\ \mathrm{cm}$-long, $30\ \mathrm{cm}$-wide and $70\ \mathrm{cm}$-high, where they form a granular column of thickness $h_d$ and solid volume fraction $\phi_{pack}$. A fluid (water, here) is then injected uniformly at a velocity $U_f$, at the base of the granular pile, passing through a porous medium (Figure \ref{fig:fluidization}). Above a threshold fluid velocity $U_{mf}$, particles become fully fluidized, so that the drag force exerted on the particles is high enough to support their weight. In this state, the mixture forms a homogeneous, stable suspension of thickness $h^{\mathrm{in}}$ and solid volume fraction $\phi_s$ which decreases as $U_f$ increases (Figure \ref{fig:fluidization}) and can be easily determined through the dilatation rate $E = \frac{h^{\mathrm{in}}}{h_d} = \frac{\phi_{pack}}{\phi_s}$ \cite{GirRis1,GirRis2,Gir_etal}.

\smallskip
Once obtained, the suspension of controlled density $\rho_m = \phi_s \rho_s + (1 - \phi_s) \rho_f$ is released down the flume of $3.5\ \mathrm{m}$-long and $30\ \mathrm{cm}$-wide, made with transparent side walls, in the manner of a dam-break. This forms a fast-moving, but short-lived, free-surface flow that sediments progressively during its travel until motion ceases. During propagation, the flow has been recorded using a semi-fast camera to capture the frontal kinematics and runouts. At the end of experiments, the deposit geometry has been carefully measured. The series of experiments presented here involves quasi-spherical, smooth glass beads (whose features are presented in Table \ref{tab:parameters}) mixed with water taken at room temperature (20$^{o} C$), of density $\rho_f = 10^3\ \mathrm{kg \cdot m^{-3}}$ and dynamic viscosity $\mu_f = 10^{-3}\ \mathrm{Pa \cdot s}$.

\begin{table}[ht]
\begin{center}
\begin{tabular}{c|c|c|c|c|c|c}
\hline 
$Materials$  & $d~(\mu \mathrm{m})$ &  $\rho_s~(\mathrm{kg \cdot m^{-3}})$  &  $\phi_{pack}$  &  $h_d~(\mathrm{cm})$  & $h^{\mathrm{in}}~(\mathrm{cm}) $ & $\phi_s/ \phi_{pack}$ \rule[5pt]{0pt}{-5pt}  \\ 
\hline 
$GB$-Water  &  $350$  &  $2500$  & $0.58$ & $19$ - $24$ & $27$  & $0.7$ - $0.9$  \rule[10pt]{0pt}{-10pt}  \\
\hline 
\end{tabular}
\end{center}
\caption{\textbf{Main experimental parameters}: Physical properties of the materials used and of suspensions generated for the dam-break flow experiments.}
\label{tab:parameters}
\end{table}

\subsection{General behavior of the flows}

\begin{figure}[h]
\centering
\includegraphics[width=0.5\linewidth]{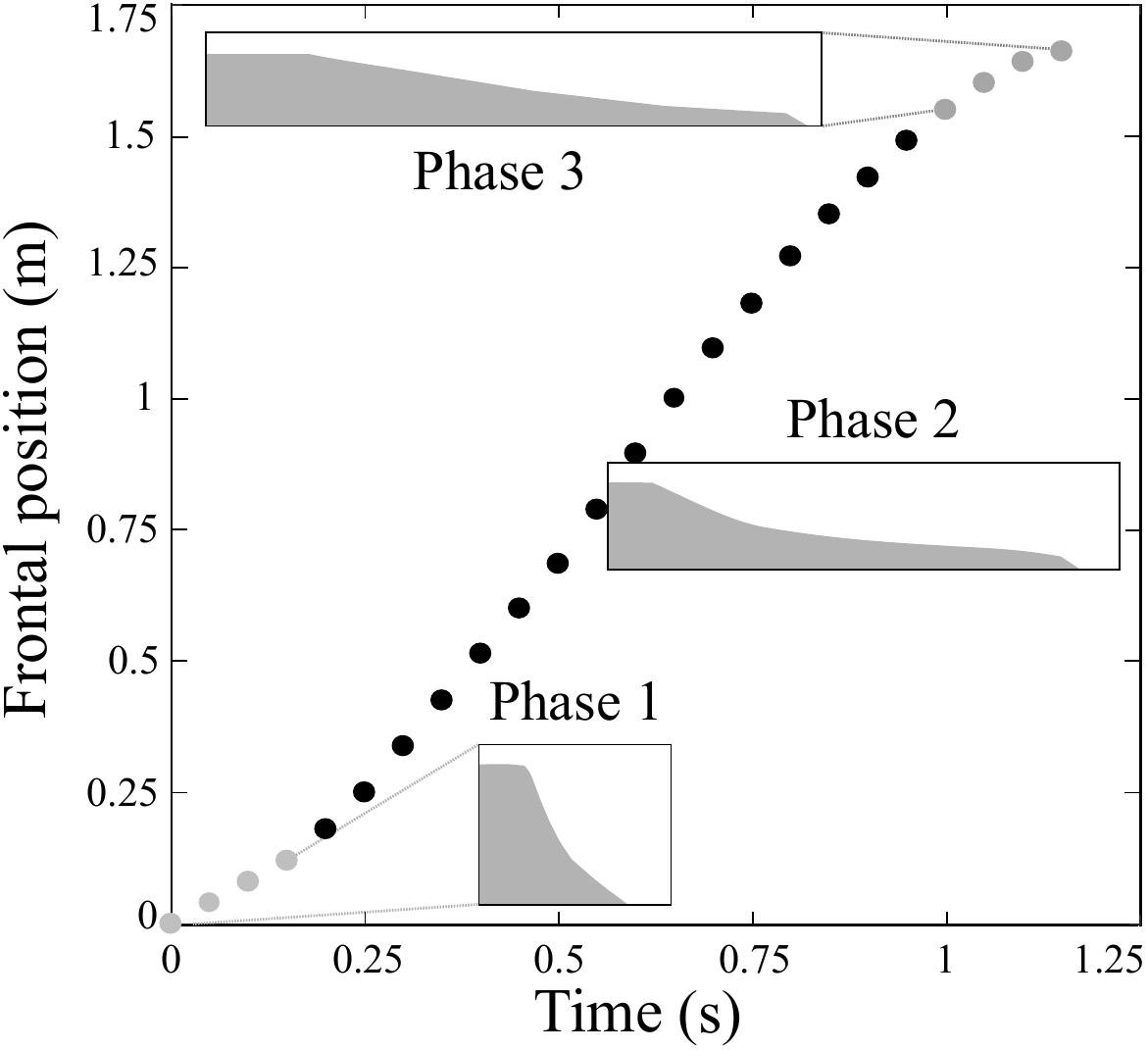}
\caption{\textbf{Kinematics profile of the flow front}: Three main phases of transport observed in experimental dam-break flows.}
\label{fig:three phases}
\end{figure}

When released, the particulate suspension travels down the flume at a mean velocity of up to $1.5\ \mathrm{m \cdot s^{-1}}$ and with a frontal speed of up to $2\ \mathrm{m \cdot s^{-1}}$. During propagation, the homogeneous mixture progressively sediments and develops an internal flow structure that can be separated into three main vertical layers: a static basal deposit that thickens progressively during the flow due to particles deposition; a flowing suspension that thins progressively by expelling water upwards and sedimenting particles downwards; an upper layer of fluid resulting from the expulsion of water. As commonly observed in classical dam-break flows, the propagation of the flow front takes place in three phases: (1) a brief gravitational collapse where no sedimentation occurs; (2) a dominant phase of constant velocity where both solid deposition and fluid expulsion develop; (3) a short decelerating phase (Figure \ref{fig:three phases}). The dam-break flow of such complex suspensions exposes a simple and systematic behavior whose dynamics is mainly controlled by the sedimentation processes, here described by the deposition velocity $U_{sed}^\sta$ that solely depends on the solid volume fractions of the suspension $\phi_s$ and of its deposit $\phi_{pack}$, as well as on the particle inertia described by the Stokes number $St_0$ \cite{GirRis1,GirRis2,Gir_etal} (Figure \ref{fig_F1-F2}). The series of experiments presented here allows us to vary $\phi_s/ \phi_{pack}$ from $0.7$ to $1$ (with a solid volume fraction $\phi_s$ ranged from $0.4$ to $0.58$) and to explore the spreading of homogeneous and stable suspensions.

\subsection{The sedimentation law}

The sedimentation processes that govern the runout of the dam-break suspension flows have been studied in details inside the reservoir, for cases where the fluid inertia is negligible compared to the viscous stresses, i.e. in the Stokes flow regime, with the aim of determining a unified sedimentation law for a wide range of particle concentrations. Different series of experiments were carried out, involving homogeneous mixtures of synthetical materials (glass and PMMA beads, FCC catalysts) or natural ones (sands, volcanic ash) characterized by a mean grain diameter ranged from $65$ to $350\ \mu \mathrm{m}$ and a solid density ranged from $1200$ to $2650\ \mathrm{kg \cdot m^{-3}}$ suspended into different types of fluids (air or water) of contrasted density (from $0.8$ to $1000\ \mathrm{kg \cdot m^{-3}}$) and viscosity (from $10^{-3}$ to $10^{-5}\ \mathrm{Pa \cdot s}$). 

\smallskip
The sedimentation velocity $U_{sed}^\sta$ was measured after cutting the fluid supply at a given mixture concentration $\phi_s/\phi_{pack}$ \cite{GirRis1}. The results analysis revealed that $U_{sed}^\sta$ is controlled by three independent parameters \cite{AmiGirRis}, being the solid volume fraction $\phi_s$, the microstructure of interstices $\phi_s/\phi_{pack}$ and a relevant Stokes number based on the Stokes velocity $St_0 = \frac{g (\rho_s - \rho_f) (\rho_s + \frac{1}{2} \rho_f) d^3}{18 ~ \mu_f^2}$, and can be described by the following general law
\begin{equation} \label{eq:U_sed}
U_{sed}^\sta = \frac{g (\rho_s - \rho_f) d^2}{18 \mu_f} (1 - \phi_s) \mathcal{F}_1\left(\frac{\phi_s}{\phi_{pack}}\right) \mathcal{F}_2 \left(St_0 \right),
\end{equation}
\begin{equation} \label{eq:F1}
\mathcal{F}_1\left(\frac{\phi_s}{\phi_{pack}}\right) = \frac{1}{\exp\left(1.9 \frac{\phi_s}{\phi_{pack}}\right) + 0.85 \left(\frac{\phi_s}{\phi_{pack}}\right)^2 \left(1-\frac{\phi_s}{\phi_{pack}}\right)^{-2/3}},
\end{equation}
\begin{equation} \label{eq:F2}
\mathcal{F}_2(St_0) = \frac{\frac{St_0}{45} + 1}{\frac{3 St_0}{45} + 1},
\end{equation}
where the two first terms $\frac{g (\rho_s - \rho_f) d^2}{18\mu_f} (1 - \phi_s)$ of equation \eqref{eq:U_sed} describe the velocity of a single, inertialess particle of density $\rho_s$ falling into a pure, viscous, infinite fluid ($\phi_s = 0$, $Re = 0$, $\frac{d}{h^{\mathrm{in}}} \ll 1$) characterized by a density $\rho_m = \phi_s \rho_s + (1 - \phi_s) \rho_f$ equivalent to that of the suspension which affects the buoyancy force acting on each particle. The third and fourth terms $\mathcal{F}_1\left(\frac{\phi_s}{\phi_{pack}}\right)$ and $\mathcal{F}_2\left(St_0\right)$ describe the effect of the mixture viscosity $\mu_m$ of the equivalent fluid which affects the drag force acting on each particle, by taking into account the mixture concentration described by $\phi_s/\phi_{pack}$ and the particle inertia described by $St_0$ \cite{AmiGirRis,Gir_etal}. 

\begin{figure}[h]
\centering
\includegraphics[width=0.5\linewidth]{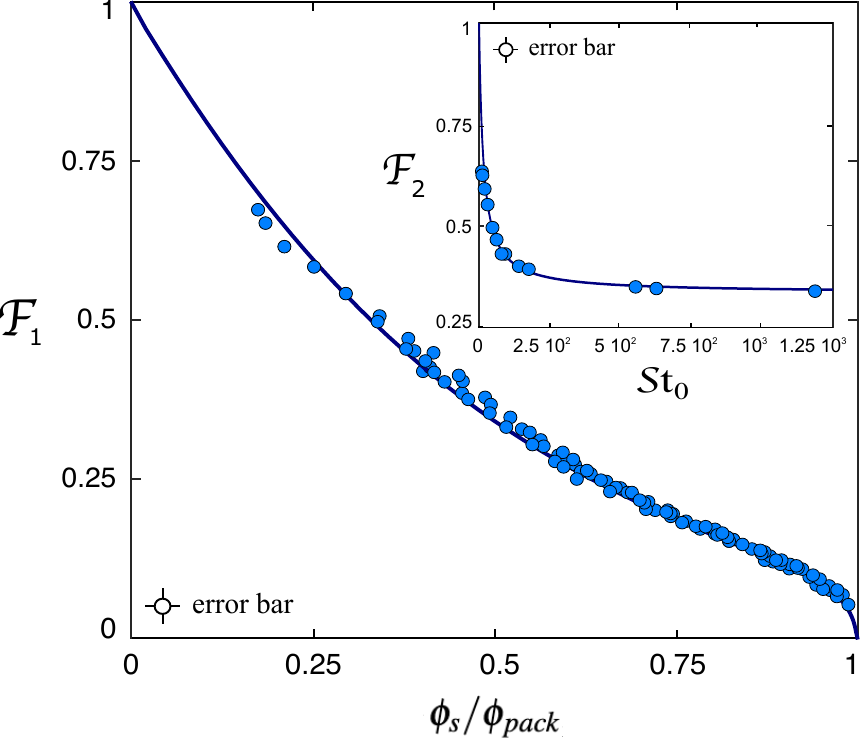}
\caption{\textbf{Shape of $\bm{\mathcal{F}_1}$ and $\bm{\mathcal{F}_2}$}: Experimental measurements exposing the decreasing functions $\mathcal{F}_1$ 
and $\mathcal{F}_2$ that respectively describe the effect of the particle concentration in the mixture and that of the particle inertia on $U_{sed}^\sta$, 
respectively expressed by \eqref{eq:F1} and \eqref{eq:F2}.}
\label{fig_F1-F2}
\end{figure}

These effects are described by independent decreasing functions (Figure \ref{fig_F1-F2}), which lead to impose two boundary conditions. In the dilute regime ($\phi_s \to 0$), equation (\ref{eq:F1}) reduces to $U_{sed}^\sta = U_0$, implying that $F_1\left(0\right) = 1$ (Figure \ref{fig_F1-F2}). Otherwise, in the dense regime ($\phi_s \to \phi_{pack}$), equation (\ref{eq:F1}) gives $U_{sed}^\sta = 0$, implying that $F_1\left(1\right) = 0$ (Figure \ref{fig_F1-F2}). Far from the packing state, $F_1\left(\frac{\phi_s}{\phi_{pack}}\right)$ is described by an exponential law, thus followed by a divergent power law close to the packing state, as commonly described in the literature \cite{Guazzelli:2018}. In between, the term proportional to $\left(\frac{\phi_s}{\phi_{pack}}\right)^2$ ensures the asymptotic matching, remaining weak close to the dilute regime while tending towards unity at the packing state \cite{Gir_etal}. Moreover, the resulting sedimentation velocity turns out to be much smaller in homogeneous suspensions than that measured in a pure fluid at rest or in heterogeneous particulate clouds which behave as clusters \cite{MetNicGua}, around which the streamlines are deflected, thus increasing the drag force exerted on individual particles. This unified law, valid for different types of non-cohesive powders and fluids, can be used to describe the sedimentation process in rapid dam-break flows of particulate suspensions in which deposition develops in the Stokes regime, despite the high Reynolds number of the flow based on the mean flow velocity \cite{GirRis2,Gir_etal}.


\section{A three-layer depth-averaged model} \label{sec:model}

In this section, we introduce the three-layer depth-averaged model developed to describe the laboratory dam-break flows. We first present the main physical assumptions stemming from the observations of the experiments, then construct the mathematical description and discuss the numerical method used to simulate the physical model, while validating this latter in the following section through a direct comparison with the measurements.

\vspace{-3mm}
\subsection{Physical assumptions}

Based on direct observations made on the general flow behavior, three main physical assumptions can be formulated in order to introduce the mathematical model.

\begin{figure}[h]
\centering
\includegraphics[width=0.7\linewidth]{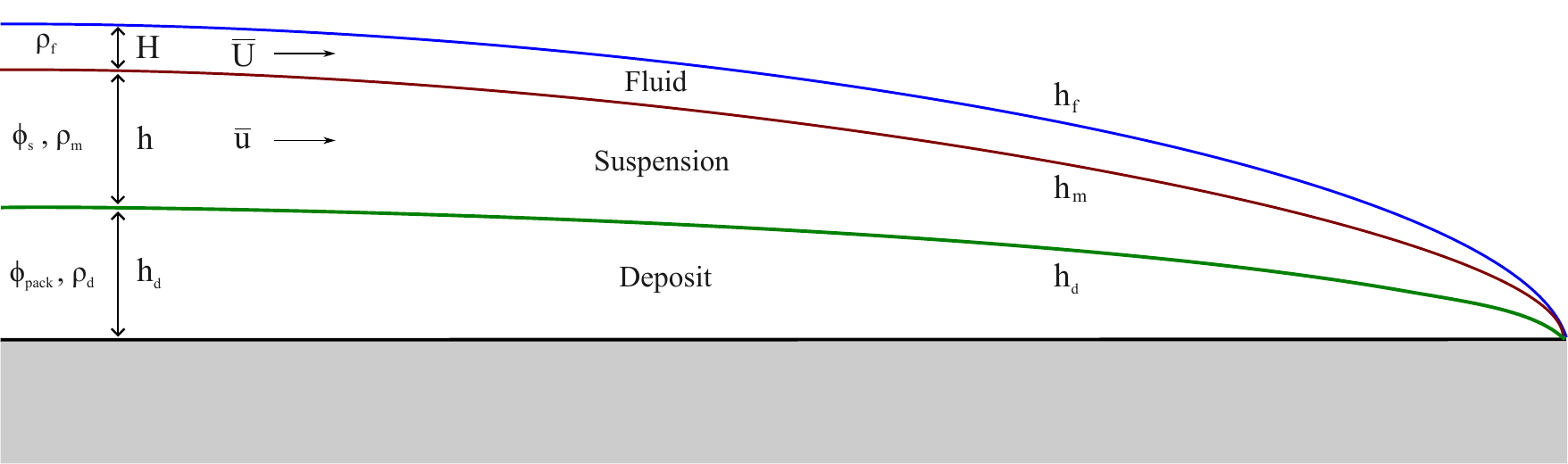}
\caption{\textbf{Internal flow structures}: Vertical stratification developed during propagation, due to the separation of the two phases and leading to the formation of three different layers (basal deposit, flowing suspension and expelled fluid).}
\label{fig: three layers}
\end{figure}

\begin{itemize}
\item[(H1)] \textbf{Structure of the dynamics.} The dynamics of the flow is characterized by three phases of transport: an initial slumping phase with no sedimentation, during which the column collapses under the influence of gravity; a dominant propagation phase where the suspension travels down the flume at constant speed, while progressively depositing particles and expelling water; a final stopping phase where the flow decelerates until motion ceases, when all the particles have settled at the bottom of the flume.

\item[(H2)] \textbf{Structure of the flow.} We assume that the particulate suspension is incompressible and develops three distinct homogeneous layers during its propagation: at the bottom, a non-moving deposit of volume fraction $\phi_{pack}$ and density $\rho_d = \rho_s \phi_{pack} + \rho_f (1-\phi_{pack})$ that grows over time due to the particles sedimentation from the above layer; a suspension of volume fraction $\phi_s$ and density $\rho_m = \rho_s \phi_s + \rho_f (1-\phi_s)$ that travels above the deposit, undergoing a basal friction proportional to $U_{sed}^\sta$, and which progressively loses particles downwards and water upwards; an upper free-surface pure fluid layer of density $\rho_f$ developed from the water expelled by the suspension, that travels without encountering any friction. The process of phase separation into these three layers is represented in Figure \ref{fig: three layers}.

\item[(H3)] \textbf{Sedimentation process.} The settling of particles develops at a rate $R_{sed}$ uniform in time and space over the moving domain of the flow, which is given explicitly by the formula
\begin{equation} \label{eq:R_sed}
R_{sed} = \frac{\rho_s}{\rho_s - \rho_f} \frac{U_{sed}^\dyn}{\frac{\phi_{pack}}{\phi_s} - 1},
\end{equation}
where the dynamic settling speed of the grains $U_{sed}^\dyn = C(Re)U_{sed}^\sta$ is proportional to the reference sedimentation velocity determined in the quasi-static configuration. In particular, the proportionality coefficient $C(Re) > 0$ depends on the flow's Reynolds number $Re = \frac{\rho_f d U_{mean}}{\mu_f}$, which is based on the mean flow velocity $U_{mean}$ measured in the experiments, the fluid properties described by $\rho_f$ and $\mu_f$, and the particle size given by the grain diameter $d$.

\end{itemize}

\subsection{Governing equations}

We develop our mathematical model under the above assumptions. Given the dimensions of the flume and the fact that the flows develop over a length much greater than their height, we consider depth-averaged equations of shallow water-type with negligible boundary effects. 

\medskip
\noindent \textbf{Modeling the slumping phase.} When the particulate suspension is released down the flume, the collapse of the initial column is essentially governed by gravitational forces. From experimental analyses, we observed that the main features of this phase corresponds to that observed in classical gravitational collapses. In particular, as the initial aspect ratio of the column $\frac{h^{\mathrm{in}}}{x^{\mathrm{in}}} > 1$ (where $h^{\mathrm{in}}$ and $x^{\mathrm{in}}$ represents the column height and length, respectively) plays a major role in the slumping dynamics, simple shallow water models fail to capture the correct acceleration of the front since the thin-layer assumption is not satisfied in this regime. In order to avoid resorting to a full Navier--Stokes formulation which is computationally more expensive, we introduce a suitable model to describe the (short) initial phase in cases where the aspect ratio is high, and then reduce the study to standard shallow water equations as soon as the flow reaches a thin-layer regime. The strategy that we follow to solve this problem has been proposed by Larrieu et al. \cite{LarStaHin}, where the collapse of a granular column is modeled as the free-falling of grains into a moving flow of small aspect ratio $\eps \ll 1$, whose dynamics can be described using shallow water equations.

\smallskip
In practice, we suppose from H1 that the suspension properties are constant in time and space, meaning that the suspension remains homogeneous with a constant volume fraction $\phi_s$ and density $\rho_m$, and that no sedimentation occurs during the initial collapse. The evolution of the suspension is thus characterized by its height $h = h(t,x)$ and by its vertically averaged lateral velocity $\bar{u} = \bar{u}(t,x)$, which both depend on time $t \in (0,t_{ff}]$ and space $x \in \R_+$. Then, the collapsing column of initial height $h^{\mathrm{in}}$ can be modeled as the evolution of a thin-layer flow of apparent initial height $h_{app} = \eps x^{\mathrm{in}}$, with $\eps \ll 1$, through the frictionless shallow water equations
\begin{equation} \label{eq:slumping-phase}
\begin{split}
& \partial_t h + \partial_x (h \bar{u}) = S_{slump}, \\[2mm]
& \partial_t (h \bar{u}) + \partial_x (h \bar{u}^2) =  S_{slump} \bar{u} - g h \partial_x h,
\end{split}
\end{equation}
where the source term $S_{slump} = S_{slump}(t,x)$ is defined as
\begin{equation} \label{eq:slumping}
S_{slump}(t,x) = g t \chi(0 < t \leq t_{ff}) \chi(0 \leq x \leq x^{\mathrm{in}}),
\end{equation}
and translates the raining of the suspension into the moving flow, over the spatial domain $x \in [0,x^{\mathrm{in}}]$ and during a time interval $t \in (0,t_{ff}]$. Here, $\chi(I)$ denotes the indicator function of the interval $I \subset \R_+$, $x^{\mathrm{in}}$ is the length of the reservoir and $t_{ff}$, defined by
\begin{equation} \label{eq:free-fall}
t_{ff} = \left( \frac{2(h^{\mathrm{in}} - h_{app})}{g} \right)^{1/2},
\end{equation}
is the typical time taken by a particle of density $\rho_m$ to fall from the true initial height $h^{\mathrm{in}}$ to the apparent height $h_{app} = \eps x^{\mathrm{in}}$.

\smallskip
\noindent \textbf{Modeling the layers interfaces.} Following hypothesis H2, as soon as the gravitational slumping is over after the critical time $t_{ff}$, the suspension propagates at constant speed down the flume and forms three distinct homogeneous layers. For any $t > t_{ff}$ and $x \in \R_+$, we denote with $h_d = h_d(t,x)$, $h_m = h_m(t,x)$ and $h_f = h_f(t,x)$, respectively the height of the deposit of solid volume fraction $\phi_{pack}$ and density $\rho_d$, the height of the homogeneous mixture of solid volume fraction $\phi_s$ and density $\rho_m$, and the height of the pure fluid of density $\rho_f$. Moreover, owing to the assumption that the incompressible flow can be described by the evolution of its longitudinal profile and that particles cease to move after having sedimented, we also need to introduce the velocity fields (horizontal and vertical velocities) of the suspension $(u,w)$ and of the fluid $(U,W)$. Here, $u = u(t,x,z)$ and $w = w(t,x,z)$ are defined for $t > t_{ff}$, $x \in \R_+$ and $h_d \leq z \leq h_m$, while $U = U(t,x,z)$ and $W = W(t,x,z)$ are defined for $t > t_{ff}$, $x \in \R_+$ and $h_m \leq z \leq h_f$. 

\smallskip
Then, the moving interfaces that characterize the evolution of each layer surface after the critical time $t_{ff}$ are modeled by the following equations
\begin{align}
& \partial_t h_f + U|_{z = h_f} \partial_x h_f - W|_{z = h_f} = 0, \label{eq:interface 1} \\[2mm]
& \partial_t h_m + u|_{z = h_m} \partial_x h_m - w|_{z = h_m} = - \left(\frac{\phi_{pack}}{\phi_s} - 1 \right) S_{sed}, \label{eq:interface 2}  \\[4mm]
& \partial_t h_d = S_{sed}, \label{eq:interface 3}
\end{align}
where the source term $S_{sed} = S_{sed}(t)$ is given by
\begin{equation*}
S_{sed}(t) = R_{sed} \chi(t > t_{ff}),
\end{equation*}
and accounts for the sedimentation of particles from the suspension over the moving domain of the flow. We stress once more that no velocity field appears in equation \eqref{eq:interface 3}, since the particles stop moving once they have settled to form the deposit. In particular, their sedimentation rate $R_{sed}$ is prescribed by hypothesis H3, using the explicit form \eqref{eq:R_sed}, and is initiated after the initial slumping time $t_{ff}$. Notice that the resulting loss of particles in the overlying suspension is taken into account by the right-hand side of \eqref{eq:interface 2}, which simply stems from a mass-conservation principle. At last, the absence of a source term in the fluid-layer equation \eqref{eq:interface 1} comes from the free-surface flow modeling.

\smallskip
\noindent \textbf{The full depth-averaged model.} We now have all the tools to introduce our model. Under the effect of gravity, the suspension and fluid layers can be described by classical incompressible Navier--Stokes and Euler equations, governing the evolution of their velocity fields $(u,w)$ and $(U,W)$ respectively. This modeling choice is a consequence of hypothesis H2, where the fluid flow is assumed frictionless over the suspension, while this latter undergoes a basal friction when flowing over the deposit. The last property is recovered by imposing a Navier-like condition at the interface with the deposit \cite{GerPer}, namely 
\begin{equation*}
\left( \kappa u - \mu_m \partial_z u \right) |_{z = h_d} = 0, \quad t > t_{ff},\ x \in \R_+,
\end{equation*}
where $\kappa$ and $\mu_m$ denote respectively the basal friction coefficient and the dynamic viscosity of the suspension. The additional boundary conditions used to close the resulting Euler--Navier--Stokes system are detailed in Appendix \ref{app:model}.

Let us now consider a thin-layer approximation of aspect ratio $\eps = \frac{h_0}{L_0}$ where $h_0$ and $L_0$ denote characteristic dimensions for the height and length of the flow. By properly rescaling the Euler--Navier--Stokes system, we look at a particular regime of low friction where $\kappa$ is taken to be $\mathcal{O}(\eps)$. Then, keeping in mind the equations for the three interfaces \eqref{eq:interface 1}--\eqref{eq:interface 3}, we apply similar arguments to those presented in the work of Gerbeau and Perthame \cite{GerPer} to derive (see Appendix \ref{app:model}), for $\eps \ll 1$, the shallow water system
\begin{align}
& \partial_t H + \partial_x (H \bar{U}) = \left(\frac{\phi_{pack}}{\phi_s} - 1 \right) S_{sed}, \label{eq:FL mass} \\
& \partial_t (H \bar{U}) + \partial_x (H \bar{U}^2) = \left(\frac{\phi_{pack}}{\phi_s} - 1 \right) S_{sed} \bar{U} \notag \\[1mm]
   &  \hspace{4.5cm} - g H \partial_x (H + h + h_d),  \label{eq:FL momentum} \\[2mm]
& \partial_t h + \partial_x (h \bar{u}) = S_{slump} - \frac{\phi_{pack}}{\phi_s} S_{sed},  \label{eq:SL mass} \\[1mm]
& \partial_t (h \bar{u}) + \partial_x (h \bar{u}^2) = \left( S_{slump} - \frac{\phi_{pack}}{\phi_s} S_{sed} \right) \bar{u} \notag \\
   &  \hspace{2.3cm} - \kappa \bar{u} \chi(t > t_{ff}) - g h \partial_x \left( \frac{\rho_f}{\rho_m} H + h + h_d \right), \label{eq:SL momentum}
\end{align}
where \eqref{eq:FL mass}--\eqref{eq:FL momentum} and \eqref{eq:SL mass}--\eqref{eq:SL momentum} express the mass and momentum conservation for the fluid and suspension layers respectively, prescribing the evolution of their flows over $t > 0$ and $x \in \R_+$ through the vertically-averaged quantities
of heights $H = H(t,x)$, $h = h(t,x)$, and mean velocities $\bar{U} = \bar{U}(t,x)$, $\bar{u} = \bar{u}(t,x)$. These are defined as $H := h_f - h_m$, $h := h_m - h_d$, and $\bar{U} := \frac{1}{H} \int_{h_m}^{h_f} U(t,x,z) \dd z$, $\bar{u} := \frac{1}{h} \int_{h_d}^{h_m} u(t,x,z) \dd z$. 

\smallskip
We first highlight that equations \eqref{eq:interface 3}--\eqref{eq:SL momentum} are compatible with the modeling of the gravitational slumping, since they reduce to equation \eqref{eq:slumping-phase} in the time interval $[0,t_{ff}]$. Indeed, the settling of particles does not occur during the initial phase since all source terms depending on $R_{sed}$ are activated after $t_{ff}$, implying that neither deposit nor fluid layers are generated during the column collapse. The sedimentation process is also mass-preserving, a property that stems directly from the definition of the layers interfaces \eqref{eq:interface 1}--\eqref{eq:interface 3}. This is not true for the conservation of momentum, where an additional basal friction $\kappa$ is imposed at the interface with the deposit. In what follows, $\kappa$ is chosen proportional to the sedimentation rate, and in particular $\kappa = R_{sed}/2$. The physical interpretation of this choice is that the loss of mass by sedimentation leads to an enhanced decrease in momentum, in line with a similar phenomenon observed for erosion processes in natural flows \cite{PudKra}. Moreover, the flow stratification into distinct sublayers is not considered here for numerical purposes, as in classical multi-layer shallow water systems \cite{Aud}, but is based on experimental observations and is characterized by well-defined physical interfaces. This feature is reflected by the term $g \frac{\rho_f}{\rho_m} h \partial_x H$ in equation \eqref{eq:SL momentum}, coming from the pressure balance at the interface between the suspension of density $\rho_m$ and the overlying fluid layer of density $\rho_f$.

\vspace{-3mm}
\subsection{Numerical strategy}

The numerical treatment of two-layer shallow water systems is rather delicate. The presence of the non-conservative products $H \partial_x h$ and $h \partial_x H$ leads to the lack of a conservation form for \eqref{eq:FL mass}--\eqref{eq:SL momentum} and to the impossibility of properly characterizing discontinuous solutions through the Rankine--Hugoniot relations. Additionally, the eigenstructure of the full two-layer system \eqref{eq:FL mass}--\eqref{eq:SL momentum} cannot be explicitly characterized and complex eigenvalues may appear. As these issues require the design of sophisticated numerical schemes to deal with the full two-layer model \cite{Par, AbgKar, CFGP, CFNPP}, it would be more desirable to treat it in an uncoupled way (for example by means of a simple implicit-explicit discretization of the non-conservative products) and to approximate equations \eqref{eq:FL mass}--\eqref{eq:FL momentum} and \eqref{eq:SL mass}--\eqref{eq:SL momentum} separately, since the eigenstructure of each layer is well-known, produces only real eigenvalues and the simulations could be performed using standard finite volume solvers for the one-layer shallow water equations, which are fast to execute. However, using a scheme for each layer independently may lead to instabilities because of the non-conservative products, and one must employ a careful decoupling strategy \cite{BouZei, BerFouMor} to successfully simulate the layers separately. Here, we shall make use of a splitting method based on a hydrostatic upwind scheme recently proposed by Berthon et al. \cite{BerFouMor}, and then combine it with the HLL solver to deal with the resulting numerical fluxes.

\smallskip
The leading idea is to reformulate equations \eqref{eq:FL mass}--\eqref{eq:SL momentum} in terms of the characteristic heights $h_1 = H + h + h_d$ and $h_2 = \frac{\rho_f}{\rho_m} H + h + h_d$, naturally arising from a computation of the steady states, and of the corresponding relative heights $X_1 = \frac{H}{h_1}$ and $X_2 = \frac{h}{h_2}$. After some simple algebraic manipulations, we obtain the transformed system
\begin{align*}
& \partial_t H + \partial_x (X_1 h_1\bar{U}) = \left(\frac{\phi_{pack}}{\phi_s} - 1 \right) S_{sed}, \\[1mm] 
& \partial_t (H\bar{U}) + \partial_x \left( X_1 \left(h_1\bar{U}^2 + \frac{g}{2} h_1^2\right) \right) = \left(\frac{\phi_{pack}}{\phi_s} - 1 \right) S_{sed} \bar{U} \\[1mm]
& \hspace{3.2cm} + \frac{g}{2} \partial_x \big( H(h_1 - H) \big) - g H \partial_x (h_1 - H),
\end{align*}
\begin{align*}
& \partial_t h + \partial_x(X_2 h_2 \bar{u}) = S_{slump} - \frac{\phi_{pack}}{\phi_s} S_{sed}, \\[1mm] 
& \partial_t (h\bar{u}) + \partial_x \left( X_2 \left(h_2\bar{u}^2 + \frac{g}{2} h_2^2\right) \right) = \left( S_{slump} - \frac{\phi_{pack}}{\phi_s} S_{sed} \right) \bar{u} \\[1mm]
& \hspace{1.1cm} - \kappa \bar{u} \chi(t > t_{ff}) + \frac{g}{2} \partial_x (h(h_2 - h)) - g h \partial_x (h_2 - h),
\end{align*}
where the coupling between layers is now taken into account implicitly through the variables $h_1$ and $h_2$. In particular, we can rewrite equations \eqref{eq:interface 3}--\eqref{eq:SL momentum} under the more convenient form
\begin{equation} \label{eq:numerical-system}
\begin{split}
& \partial_t w_1 + \partial_x \big( X_1 f(W_1) \big) = P(H, h_1) + S_1(\bar{U}), \\[2mm]
& \partial_t w_2 + \partial_x \big( X_2 f(W_2) \big) = P(h, h_2) + S_2(\bar{u}), \\[2mm]
& \partial_t h_d = S_{sed},
\end{split}
\end{equation}
by using the notations $w_1 = (H, H \bar{U})^\intercal$ and $w_2 = (h, h \bar{u})^\intercal$ for the original shallow water variables, $W_1 = (h_1, h_1 \bar{U})^\intercal$ and $W_2 = (h_2, h_2 \bar{u})^\intercal$ for the new characteristic variables, $f(W_1) = \left( h_1 \bar{U}, h_1 \bar{U}^2 + \frac{g}{2} h_1^2 \right)^\intercal$ and $f(W_2) = \left( h_2 \bar{u}, h_2 \bar{u}^2 + \frac{g}{2} h_2^2 \right)^\intercal$ for the fluxes, $P(H,h_1) = \left( 0, \frac{g}{2} \partial_x \big( H(h_1 - H) \big) - g H \partial_x (h_1 - H) \right)^\intercal$ and $P(h,h_2) = \left( 0, \frac{g}{2} \partial_x \big( h(h_2 - h) \big) - g h \partial_x (h_2 - h) \right)^\intercal$ for the pressure terms (notice that the structure of $f$ and $P$ is the same for the two layers, hence the notation with no dependence on the indices) and finally
\begin{equation*}
\begin{split}
& S_1 (\bar{U}) = \left( \begin{array}{c} \left(\frac{\phi_{pack}}{\phi_s} - 1 \right) S_{sed} \\[3mm]  \left(\frac{\phi_{pack}}{\phi_s} - 1 \right) S_{sed} \bar{U} \end{array} \right), \\[4mm]
& S_2 (\bar{u}) = \left( \begin{array}{c} S_{slump} - \frac{\phi_{pack}}{\phi_s} S_{sed} \\[3mm]   \left( S_{slump} - \frac{\phi_{pack}}{\phi_s} S_{sed} \right) \bar{u} - \kappa \bar{u} \chi(t > t_{ff}) \end{array} \right),
\end{split}
\end{equation*}
for the source terms involving the slumping process, the sedimentation of particles from the suspension and the friction. 

\smallskip
The aim is to design a suitable finite volume scheme to approximate equation \eqref{eq:numerical-system}. For this, we introduce a time step $\Delta t > 0$ and a space step $\Delta x > 0$, and define corresponding uniform meshes $(t^n)_{n \in \mathbb{N}}$ and $(x_{i+\frac{1}{2}})_{i \in \mathbb{N}}$ of $\R_+$, such that $t^n = n \Delta t$ and $x_{i + \frac{1}{2}} = \left( i + \frac{1}{2} \right) \Delta x$. Then, at each time $t^n$ and on each spatial cell $(x_{i-\frac{1}{2}}, x_{i+\frac{1}{2}})$, we consider constant approximations of the original variables $w_{1,i}^n = (H_i^n, H_i^n \bar{U}_i^n)^\intercal$ and $w_{2,i}^n = (h_i^n, h_i^n \bar{u}_i^n)^\intercal$ and of the new ones, as
\begin{equation*}
\begin{split}
& h_{1,i}^n = H_i^n + h_i^n + h_{d,i}^n \quad \textrm{with} \quad X_{1,i}^n = \frac{H_i^n}{h_{1,i}^n}, \\[2mm]
& h_{2,i}^n = \frac{\rho_f}{\rho_m} H_i^n + h_i^n + h_{d,i}^n \quad \textrm{with} \quad X_{2,i}^n = \frac{h_i^n}{h_{2,i}^n}.
\end{split}
\end{equation*}
Moreover, we also introduce suitable discretizations for the following terms
\begin{equation*}
\begin{split}
& S_{slump,i}^n = g t^n \chi({0 \leq x_i \leq x_0}) \chi({0 < t^n \leq t_{ff}}), \\[2mm]
& S_{sed}^n = R_{sed} \chi{t^n > t_{ff}}, 
\end{split}
\end{equation*}
which are activated respectively during the initial slumping phase and during the following sedimentation-driven phase.

\smallskip
From here, the updated states $w_{1,i}^{n+1}$ and $w_{2,i}^{n+1}$ are determined through a two-step scheme in time. We first take into account the fluxes $f$ and the pressure terms $P$ to compute intermediate values $w_{1,i}^*$ and $w_{2,i}^*$ from the initial conditions $(w_{j,i}^n)_{j=1,2}$ \cite{BerFouMor}, namely
\begin{equation} \label{eq:numerical-scheme}
\begin{split}
& w_{j,i}^* = w_{j,i}^n - \frac{\Delta t}{\Delta x} \Big( X_{j,i+\frac{1}{2}}^n f_{\Delta x}(W_{j,i}^n,W_{j,i+1}^n) \\[2mm]
&  \hspace{3cm} - X_{j,i-\frac{1}{2}}^n f_{\Delta x}(W_{j,i-1}^n,W_{j,i}^n) \Big) \\[2mm]
& \hspace{1.5cm} + \frac{g}{2} \frac{\Delta t}{\Delta x} \left( \begin{array}{c}  0  \\[2mm]  h_{j,i-\frac{1}{2}}^n h_{j,i+\frac{1}{2}}^n (X_{j,i+\frac{1}{2}}^n - X_{j,i-\frac{1}{2}}^n)  \end{array} \right),
\end{split}
\end{equation}
for $j = 1,2$, where the classical shallow water numerical fluxes $f_{\Delta x} = (f_{\Delta_x}^h, f_{\Delta x}^{hu})^\intercal$ are approximated using the HLL solver \cite{Bou}, while the values $(h_{j,i+\frac{1}{2}}^n)_{i \in \mathbb{N}}$ and $(X_{j,i+\frac{1}{2}}^n)_{i \in \mathbb{N}}$, $j = 1,2$, at the interfaces of the spatial grid are defined through an upwind method, having the form
\begin{equation*}
\begin{split}
& h_{j,i+\frac{1}{2}}^n = \left\{ \begin{array}{ll} W_{j,i}^n & \quad \textrm{if} \ f_{\Delta x}^h(W_{j,i}^n, W_{j,i+1}^n) > 0, \\[2mm]
W_{j,i+1}^n & \quad \textrm{elsewhere}, \end{array} \right.  \qquad  j = 1,2, \\[4mm]
& X_{1,i+\frac{1}{2}}^n = \left\{ \begin{array}{ll} \frac{H_i^n}{h_{1,i}^n} & \quad \textrm{if} \ f_{\Delta x}^h(W_{1,i}^n, W_{1,i+1}^n) > 0, \\[2mm]
\frac{H_{i+1}^n}{h_{1,i+1}^n} & \quad \textrm{elsewhere}, \end{array} \right. \\[4mm]
& X_{2,i+\frac{1}{2}}^n = \left\{ \begin{array}{ll} \frac{h_i^n}{h_{2,i}^n} & \quad \textrm{if} \ f_{\Delta x}^h(W_{2,i}^n, W_{2,i+1}^n) > 0, \\[2mm]
\frac{h_{i+1}^n}{h_{2,i+1}^n} & \quad \textrm{elsewhere}. \end{array} \right.
\end{split}
\end{equation*}
In order to avoid possible instabilities and to ensure that the scheme preserves the nonnegativity of the layers' heights \cite{BerFouMor}, we also impose here the CFL-like conditions
\begin{equation*}
\begin{split}
& \frac{\Delta t}{\Delta x} \ \max\limits_{\substack{j = 1,2 \\ i \in \mathbb{N}}} \ \left| \lambda^{\pm}(W_{j,i}^n,W_{j,i+1}^n) \right| \leq \frac{1}{2}, \\[2mm]
& \frac{\Delta t}{\Delta x} \Big( \max \big(0, f_{\Delta x}(W_{j,i}^n,W_{j,i+1}^n) \big) \\[0.5mm]
&  \hspace{1.8cm} - \min \big(0, f_{\Delta x}(W_{j,i-1}^n,W_{j,i}^n) \big) \Big) \leq h_{j,i}^n, \quad j = 1,2,
\end{split}
\end{equation*}
where $\lambda^{\pm}(W_{j,i}^n,W_{j,i+1}^n)$ denote some numerical approximations of the external eigenvalues of system \eqref{eq:numerical-system}, associated with the numerical flux $f_{\Delta x}$. 

\smallskip
Starting from the initial conditions $(w_{j,i}^*)_{j=1,2}$, we then consider an explicit discretization of the source terms $S_1$ and $S_2$, to finally recover
\begin{equation*}
\begin{split}
& w_{1,i}^{n+1} = w_{1,i}^* + \Delta t \left( \begin{array}{c} \left(\frac{\phi_{pack}}{\phi_s} - 1 \right) S_{sed}^n \\[3mm]  \left(\frac{\phi_{pack}}{\phi_s} - 1 \right) S_{sed}^n \bar{U}_i^* \end{array} \right), \\[4mm]
& w_{2,i}^{n+1} = w_{2,i}^* + \Delta t \left( \begin{array}{c} S_{slump,i}^n - \frac{\phi_{pack}}{\phi_s} S_{sed}^n \\[3mm]   \left( S_{slump,i}^n - \frac{\phi_{pack}}{\phi_s} S_{sed}^n \right) \bar{u}_i^* - \kappa \bar{u}_i^* \chi(t^n > t_{ff}) \end{array} \right), \\[8mm]
& h_{d,i}^{n+1} = h_{d,i}^n + \Delta t S_{sed}^n.
\end{split}
\end{equation*}
To provide some important remarks, we highlight that the numerical scheme \eqref{eq:numerical-scheme} is well-balanced and preserves the nonnegativity of the layers' heights $h$ and $H$ (and also $h_d$, obviously). The discretization of the sedimenting process then guarantees that we also have preservation of the total suspension mass. In particular, we perform the last numerical update giving $(w_{1,i}^{n+1}, w_{2,i}^{n+1}, h_{d,i}^{n+1})$, where we account for the sedimentation of particles, as long as there is enough mass to add to the deposit. As a direct consequence, the flow naturally stops once all the particles inside the suspension have settled at the bottom of the channel. We also point out that the scheme is designed up to a topography translation, therefore we need to impose over the whole domain of computation that $h_1 > \delta$ and $h_2 > \delta$, where $\delta > 0$ is an arbitrary small constant. At last, while in this work we only consider a first-order method in time and space (which is enough for these simulations), we mention that more robust MUSCL schemes combined with Heun's method could be implemented to achieve second-order accuracy \cite{BerFouMor}.

\begin{figure*}
\centering
\includegraphics[width=0.9\linewidth]{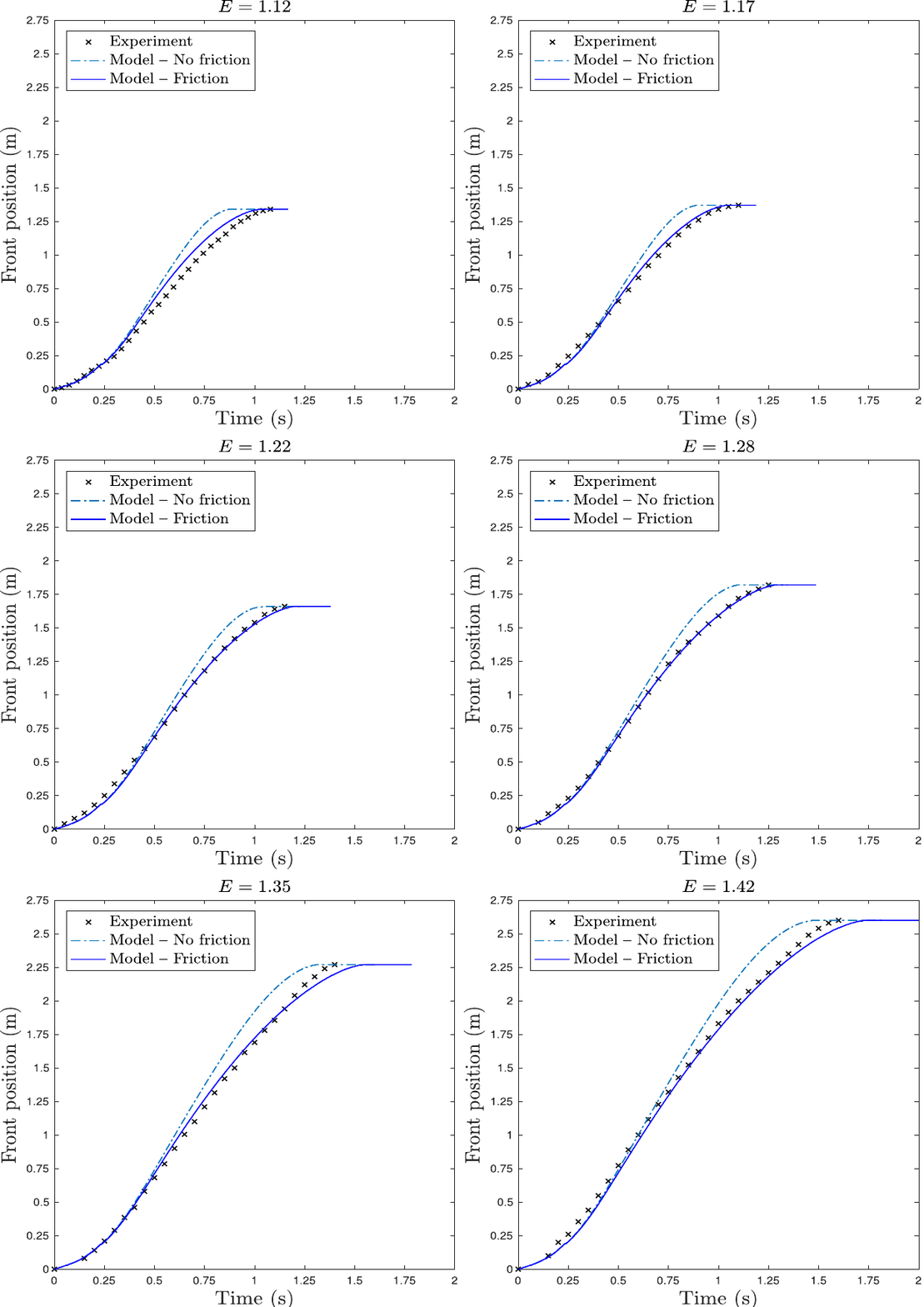}
\caption{\textbf{Flow kinematics}: Temporal evolution of the frontal position for different initial expansions, from $12 \%$ to $42 \%$. Experimental results are represented by black crosses, numerical predictions with no basal friction are represented in light blue dashed lines, while those with an evolving basal friction given by $\kappa = R_{sed}/2$ are represented in dark blue continuous lines.}
\label{fig:kinematic profiles}
\end{figure*}

\section{Model predictions of dam-break sedimenting suspensions} \label{sec:model-validation}



We investigate a series of experiments involving particulate suspensions generated in a reservoir by fluidizing a packed bed of volume fraction $\phi_{pack}$. Each suspension can be characterized by a specific solid volume fraction $\phi_s$, through the expansion rate $E = \frac{\phi_{pack}}{\phi_s}$ \cite{Gir_etal}. We focus our attention on the fully fluidized regime, where the fluid-particles mixture can be considered as homogeneous. Specifically, interaction forces between particles are negligible and water flows through interstices in a laminar way, so that no clustering of particles is observed, and the suspension can be modeled as an equivalent fluid of density $\rho_m = \phi_s \rho_s + (1-\phi_s) \rho_f$, viscosity $\mu_m = \frac{\mu_f }{\mathcal{F}_1\left(\frac{\phi_{pack}}{\phi_s}\right) \mathcal{F}_2\left(St_0\right)}$ \cite{AmiGirRis}, and volume fraction $\phi_s$. This particular regime is achieved for $E \geq 1.12$, which corresponds to a volume expansion of up to $12 \%$. We consider six experiments, distinguished by six different values of $E$ ranging from $1.12$ up to $1.42$, for a maximum volume expansion of $42 \%$. When the suspension is released down the flume, the mixture behaves as a classical dam-break flow. In this study, we aim at capturing two of their main features: the kinematics of the flow through the spatio-temporal evolution of the frontal position, and the deposits geometry.

\subsection{Kinematics of the flow fronts}

We showed in Figure \ref{fig:three phases} the typical kinematics curve of a dam-break flow, including its three main phases of transport. This characteristic profile is also shown in Figure \ref{fig:kinematic profiles} where we depict, for each of the six considered expansions, the evolution in time of the frontal position of the experimental flows (black crosses). For increasing expansions (from $E = 1.12$ to $E = 1.42$), the traveled distance of the flows significantly increases from about $1.25\ \mathrm{m}$ to more than $2.5\ \mathrm{m}$, while their duration increases from about $1.25\ \mathrm{s}$ to $2\ \mathrm{s}$. Furthermore, the profiles exhibit increasing frontal velocities during propagation, that can be better modeled with equations \eqref{eq:interface 3}--\eqref{eq:SL momentum} (dark blue line). 

\smallskip
\noindent \textbf{The gravitational collapse.} As the thin-layer approximation is not valid during the initial collapse, since the column height ($h{^\mathrm{in}} = 0.27\ \mathrm{m}$) is much larger than its length ($x^{\mathrm{in}} = 0.105\ \mathrm{m}$), we transferred the approach of Larrieu et al. \cite{LarStaHin} to our setting by letting the suspension falling into an apparent flowing layer of small height $x^{\mathrm{in}} / 10$, at speed $(2gh)^{1/2}$ and during a time interval $[0,t_{ff}]$, where $t_{ff} \approx 0.23 s$ denotes the representative free-fall time of a particle, which is encoded in the model through the source terms depending on $S_{slump}$ and is defined by \eqref{eq:slumping}. From Figure \ref{fig:kinematic profiles} we can see that this strategy allows us to well reproduce the accelerating behavior in each experiment and to capture, in particular, the correct initial frontal velocity at the beginning of the second phase of transport.

\smallskip
\noindent \textbf{The sedimentation-dominated dynamics.} A progressive sedimentation takes place in the following dominant phase of transport, where the suspension travels down the flume at constant speed. Despite the high Reynolds number of the flows, the sedimentation of particles may develop independently in the Stokes flow regime where the effects of the fluid inertia are negligible, so that their settling speed remains similar to the one measured in the static configuration \cite{GirRis2, Gir_etal}. This means that taking $U_{sed}^\sta$, described in  \eqref{eq:U_sed}, as the reference settling velocity represents a reasonable simplification. In practice, we expect that the flow agitation could slightly influence this reference velocity, leading to the modeling assumption on the form of the sedimentation rate $R_{sed}$, given by equation \eqref{eq:R_sed} 
as a function of $Re$. This parameter $R_{sed}$ governs the runout distance of the flow modeled by \eqref{eq:interface 3}--\eqref{eq:SL momentum}, and decreases with increasing expansions: the more the suspension is fluidized, the more agitation affects the vertical motion of the particles, the more their sedimentation is delayed (see Figure \ref{fig:fitting}) and thus the distance traveled by the flow is extended. Such feature is encoded here by the coefficient $C(Re)$, which encapsulates the effects of agitation within the suspension depending on the flow's specific Reynolds number defined by $Re = \frac{\rho_f d U_{mean}}{\mu_f}$ in terms of the mean flow velocity $U_{mean}$ measured in the experiments. Using $C(Re)$ as a free parameter in our model allows us to fit the runout distance of each experiment. A comparison between the kinematics profiles measured from the experiments (black crosses) and those predicted numerically is exposed in Figure \ref{fig:kinematic profiles} in cases where a basal friction at the interface with the deposit is included (dark blue continuous lines) or disregarded (light blue dashed lines). In particular, we fix $\kappa = R_{sed}/2$. We can notice that the model with $\kappa = 0$ always underestimates the slope of the experimental profiles, justifying the use of friction to recover the correct speed of propagation during this phase. This is confirmed by the good approximations provided by the curves in dark blue.

\begin{figure}[h]
\centering
\includegraphics[width=0.5\linewidth]{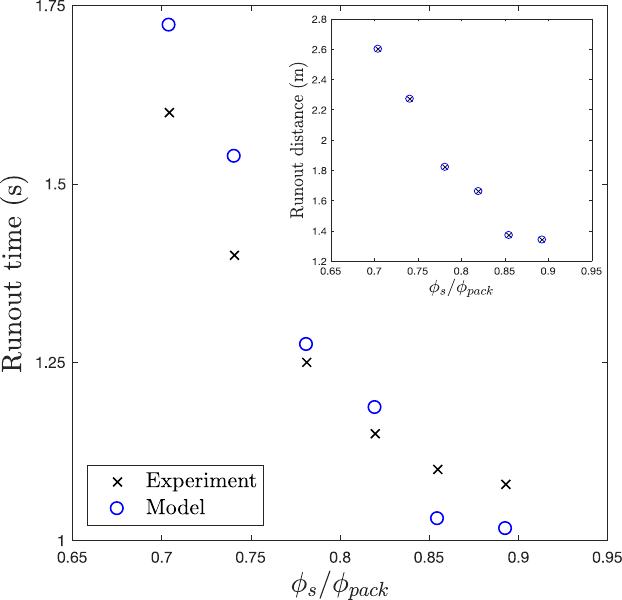}
\caption{\textbf{Runout time and distance}: Duration and traveled distance of experimental flows (black crosses) as functions of the mixture concentration, as well as predictions gained from equations \eqref{eq:interface 3}--\eqref{eq:SL momentum} by taking $\kappa = R_{sed}/2$ (blue circles).}
\label{fig:runouts}
\end{figure}

\smallskip
\noindent \textbf{The stopping phase.} During the last phase, the suspension experiences a deceleration and motion ceases once all particles have settled, forming the final deposit. We observe in Figure \ref{fig:kinematic profiles} that all the numerical profiles display a horizontal behavior at the end of the propagation. This is related with the fact that once the front has stopped moving, the suspension located at the rear of the flow continues sedimenting on top of the deposit and expelling water upward, as prescribed by the source terms that depend on $S_{sed}$ in equations \eqref{eq:interface 3}--\eqref{eq:SL momentum}. Conversely, the stopping phase of the experimental flows is characterized by two successive waves of motion: a first one corresponding to the propagation of the suspension which lasts until the front is at rest, and a second one associated with the remobilization of a small volume of suspension at the rear of the deposit, forming a secondary front that transports the material forwards and extends the runout length of up to $8 \%$ \cite{GirDruRoc}. We speculate that this phenomenon is related to surface instabilities and is caused by a late fluidization (due to a deposit compaction that creates a pressure release) developing at the rear part, which leads to a remobilization of a small volume of material and to its subsequent transport over the deposit. In particular, the second wave is observed to be more significant for greater expansions, but its solid concentration progressively decreases with $E$. Since the present model is not designed to capture such secondary effects, we have chosen to reproduce the experimental kinematic profiles only up to the end of the first wave (meaning that we do not take into account the second wave in the experimental curves of Figure \ref{fig:kinematic profiles}). The drawback of this choice is a loss of precision in the model predictions for the flow duration, as shown by Figure~\ref{fig:runouts}. While the runout distances of the experimental flows are satisfyingly captured by the model, we can observe a discrepancy between the experimental and the predicted runout times, which correspond to the end of the first wave. This might be explained by the fact that we are not taking into account in equations \eqref{eq:interface 3}--\eqref{eq:SL momentum} essential information about the dynamics of the second wave, making it difficult to identify the correct physical runout time and compare it with the numerical one.


\begin{figure}[h]
\centering
\includegraphics[width=0.5\linewidth]{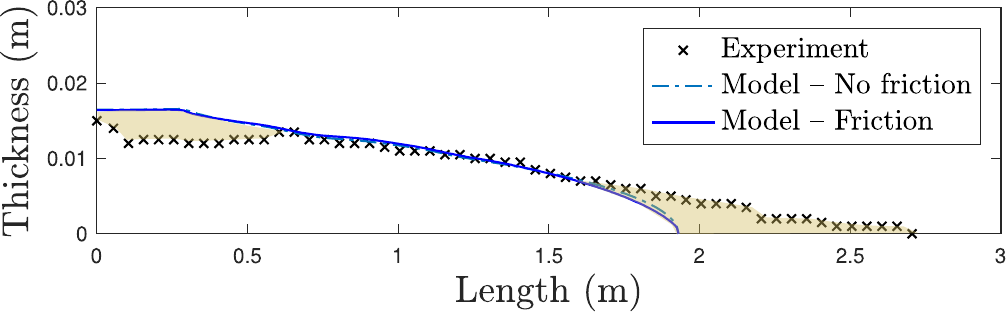}
\caption{\textbf{Second wave}: Visualization of the second wave through its impact on the deposit geometry for $E = 1.28$. The predicted excess of particles in the experimental deposit, represented in back, is comparable to that missing at the front (see brown colored areas).}
\label{fig:secondary wave}
\end{figure}

\subsection{Geometry of the final deposits}

The second wave can be visualized a posteriori by analyzing the deposits geometry and comparing them with the numerical predictions. The case of a volume expansion $E = 1.28$ offers the best example. In Figure \ref{fig:secondary wave}, we show the final deposit measured experimentally (black crosses) and that obtained numerically from equations \eqref{eq:interface 3}--\eqref{eq:SL momentum} (dark/light blue lines, with or without friction), highlighting in light brown the areas of the volume transported by the second wave. We notice that the model fails to correctly reproduce the experimental deposit both at the rear and at the front, but the areas appear to compensate, so that the discrepancy may correspond to the remobilized material through the second wave (which is not accounted by the model). The latter flows over the deposit and transports particles forwards, explaining the differences between the predicted deposit geometry and the experimental one.

\begin{figure*}[h]
\centering
\includegraphics[width=0.95\linewidth]{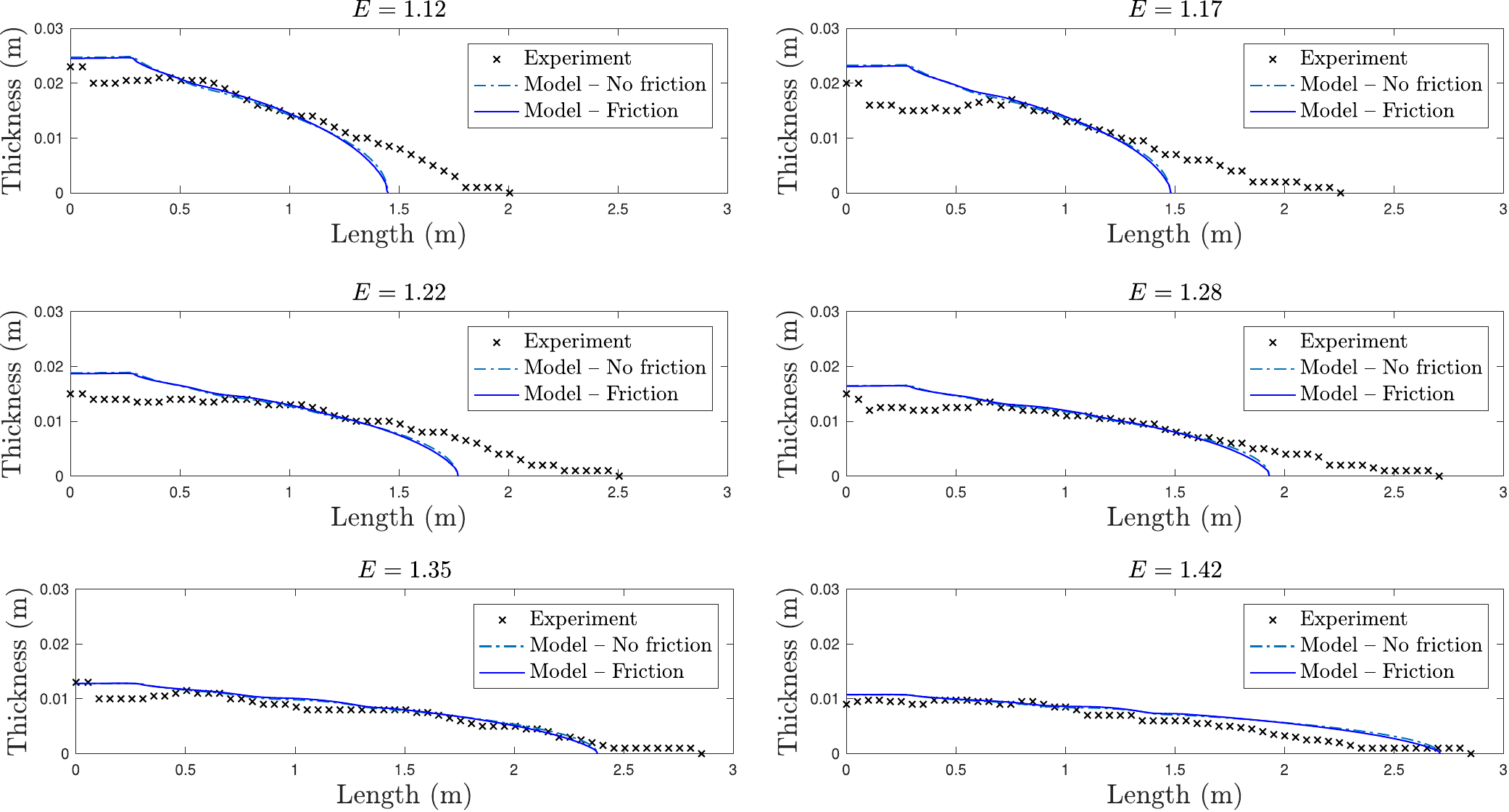}
\caption{\textbf{Final deposits}: Geometry of the final deposits measured at the end of the experiments (black crosses). Comparison with the shapes predicted by the model, with or without basal friction (dark blue continuous line/light blue dashed line).}
\label{fig:deposits}
\end{figure*}

We finally display in Figure \ref{fig:deposits} a similar comparison between experimental and numerical deposits, for all the considered values of $E$. As the expansion grows, the geometry of the measured deposits becomes more elongated and exhibits a more gentle slope, as well as a lower maximum thickness. This is due to the decrease of the sedimentation rate which acts in promoting the transport of particles down the flume. We can also observe how the predictions from equations \eqref{eq:interface 3}--\eqref{eq:SL momentum} provide better results for higher expansions, when the volume of particles remobilized by the second wave becomes negligible. In any case, this second wave does not modify significantly the shape of the primary deposits, leading to global relevant numerical predictions.

\subsection{Discussion}

We discuss here the main investigations that we plan to carry out in the near future to adjust and improve our model. 

\begin{figure*}
\includegraphics[width=0.7\linewidth]{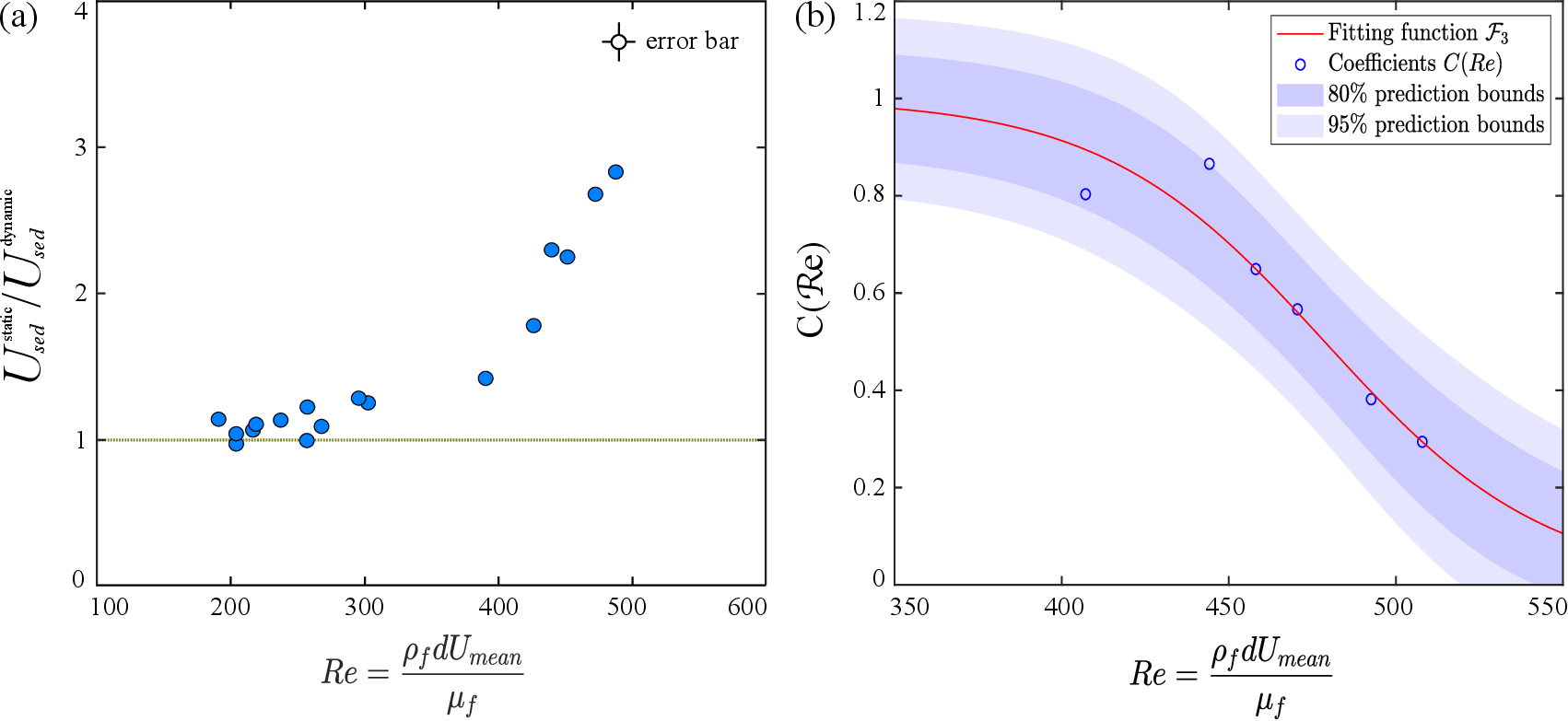}
\caption{\textbf{Threshold $Re$ number}: (a) Experimental measurements of the non-dimensional ratio $\displaystyle U_{sed}^\sta/U_{sed}^\dyn$ as a function of the Reynolds number $Re$ that compares the fluid inertia to the viscous stress at the scale of the particles; (b) Coefficient $C(Re)$ which describes the delay of particles sedimentation velocity as a function of $Re$, i.e. due to the flow agitation. The fitting function \eqref{eq:F3} is plotted along the $80\%-95\%$ prediction bounds.}
\label{fig:fitting}
\end{figure*}

\smallskip
\noindent \textbf{Static and dynamic sedimentation velocities.} We present a general law \eqref{eq:U_sed} which describes the sedimentation velocity $U_{sed}^\sta$ of fine, non-cohesive powders in a low-inertia fluid in static suspensions \cite{GirRis1, AmiGirRis, Gir_etal}. Thus, we focus our attention on the sedimentation process developed in flowing suspensions. Comparisons between $U_{sed}^\sta$ and $U_{sed}^\dyn$ highlight that as long as $Re \leq 250$, the sedimentation velocity measured in flowing suspensions $U_{sed}^\dyn$ is of the same order of magnitude than that measured in static suspensions (Figure \ref{fig:fitting}a), highlighting that no significant vertical velocity fluctuations can alter the particles motion, so that particles settle vertically throughout the flow and allow us to decouple this process from that of the flow and remains quasi-identical to that studied in detail in static suspensions. Above the threshold value of $Re_c > 250$, the fluid inertia becomes sufficiently important to develop fluctuating eddies able to significantly disturb and delay their sedimentation. This effect is taken into account in our model through the coefficients $C(Re)$, that can be approximated by a function $\mathcal{F}_3(Re)$ of the form
\begin{equation} \label{eq:F3}
\mathcal{F}_3(x) = \frac{0.53}{0.53 + \exp\big(0.06 (x - 2Re_c)\big)},
\end{equation}
which decreases with $Re$ and thus with decreasing mixture concentrations (Figure \ref{fig:fitting}b). This means that, when $Re \leq Re_c$, the suspension dynamics can be described as a low-viscosity quasi-parallel flow \cite{GirRis2} where particles settle vertically at the velocity $U_{sed}^\sta$, thus taking $C(Re) \approx 1$. Otherwise, when $Re \geq Re_c$, the suspension becomes agitated which delay the sedimentation process whose description requires an additional effect $C(Re)$. A more detail experimental investigation is required to analyze more deeply this aspect.

\smallskip
\noindent \textbf{Modeling the second wave.} The model is presently able to reproduce relevant flow front kinematics of fully-fluidized suspensions, as well as their final deposits. It appears however that the model loses some precision when predicting the flow duration and deposits at low expansion rates, as shown in Figure \ref{fig:kinematic profiles} and Figure \ref{fig:deposits} respectively. This can be explained by the presence of a secondary wave of flowing material, observed experimentally, that arises at the surface of the main deposit. Its origin can be explained by the fluid expel during the deposit compaction, at the rear of the flow, which can lead to its partial fluidization and remobilization, such as elongating the runout distance, thus modifying the deposit morphology. Modeling this phenomena may require a multi-phase approach which is beyond the scope of this paper.

\section{Conclusions} \label{sec:conclusions}

In this work, we conducted laboratory experiments of the dam-break flows of homogeneous suspensions of different concentrations, made of fine, non-cohesive powders and water, whose dynamics is mostly dominated by sedimentation. Starting from a coupled Euler--Navier--Stokes formulation, we derived a simple three-layer depth-integrated shallow water model to describe the flow evolution and deposits. The resulting system of equations includes three source terms built to model respectively the column collapse (during the initial non-shallow water phase), the propagation velocity of the suspension front and the sedimentation process that forms the basal deposit. The first term is obtained through an adaptation of the method from the work of Larrieu et al. \cite{LarStaHin} designed for granular collapses to describe the gravitational slumping of the suspension via the free-falling of its particles into a moving apparent thin layer. The second one is given by a small basal friction (chosen proportional to the particle settling velocity, as done in erosion processes \cite{PudKra}) experienced by the suspension layer when traveling over the deposit. The last one translates the particles sedimentation rate in flowing suspensions, inferred from experiments and the physical model. The next step will consist in including the effect of the fluid inertia, described by the Reynolds number, in the general sedimentation law. The equations were solved numerically using a splitting strategy based on a hydrostatic upwind scheme that allows to treat the system in an uncoupled way \cite{BerFouMor}, while the evolution of each layer is separately approximated through a finite volume scheme using the HLL solver to discretize the numerical fluxes. The proposed model allows to recover the first order frontal kinematic profiles as well as the deposits geometry, and to capture in particular the relevant flows front velocity after the initial slumping phase. The development of a second wave at the end of the primary flows motion prevents the accurate reproduction of the deposits geometry for lower values of the expansion. A possible way to incorporate this effect may be to consider a more elaborated two-phase flow formulation with a time-evolving volume fraction $\phi_s$ \cite{MerTamRoc, BFMN2, BauKam}. These investigations are left for future developments.


\section*{Acknowledgments}
This study was supported by the Region Centre-Val de Loire through the Contribution of the ``Academic Initiative'' RHEFLEXES/2019-00134935. A.B. also wishes to acknowledge supports from the GNFM group of INdAM (Italian National Institute of High Mathematics), from the Austrian Science Fund (FWF), within the Lise Meitner project No. M-3007 (Asymptotic Derivation of Diffusion Models for Mixtures), and from the European Union’s Horizon Europe research and innovation programme, under the Marie Sklodowska-Curie grant agreement No. 101110920 (MesoCroMo - A Mesoscopic approach to Cross-diffusion Modelling in population dynamics).


\appendix

\section{Derivation of the interfaces' equations} \label{app:interfaces}

\begin{figure}[h]
\centering
\includegraphics[width=0.5\linewidth]{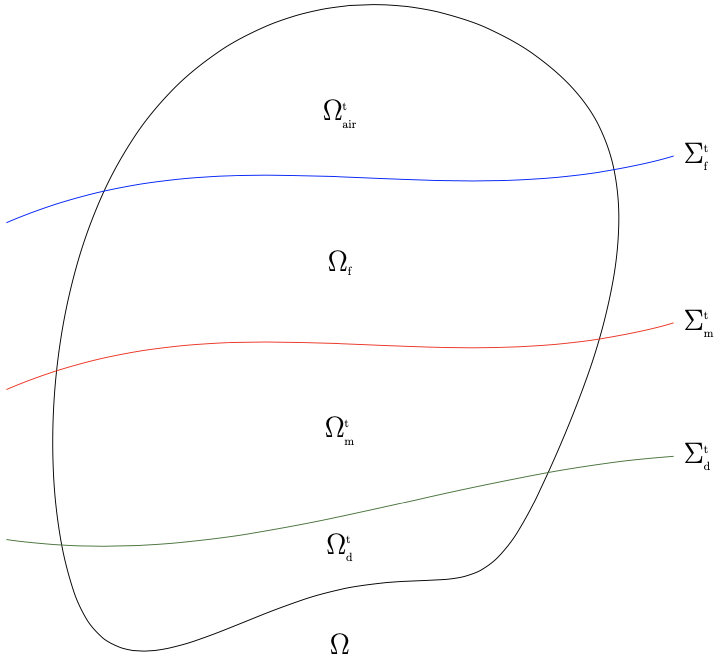}
\caption{\textbf{Control volume}: Sketch of the control volume $\Omega^t$ used to determine the evolution equations for the interfaces.}
\label{fig:control volume}
\end{figure}

In this first appendix, we focus our attention on the derivation of equations \eqref{eq:interface 1}--\eqref{eq:interface 3}, describing the evolution of the interfaces that separates the three layers of the flow. Let us take $t > t_{ff}$, where $t_{ff}$ is the critical time \eqref{eq:free-fall} starting from which these three interfaces form. We begin by considering a fixed control volume $\Omega$ around the three interfaces, as well as four mobile domains that we call $\Omega_d^t$ for the deposit, $\Omega_m^t$ for the grains--water mixture, $\Omega_f^t$ for the pure fluid and $\Omega_{\textrm{air}}^t$ for the upper ambient air (see Figure \ref{fig:control volume}). Then, assuming incompressibility holds, we may write the continuity equation for the two-dimensional flow inside each layer in the general non-conservative form
\begin{equation} \label{eq:continuity-equation}
\partial_t \rho + \rho \partial_x u + \rho \partial_z w = 0,
\end{equation}
where the inhomogeneity of the medium is encoded by a piecewise constant (constant in each of the four domains $\Omega_d^t$, $\Omega_m^t$, $\Omega_f^t$ and $\Omega_\textrm{air}^t$) density $\rho$ and by a piecewise regular (similarly, regular in each of the four domains) velocity field $v = \left( \begin{array}{c} u \\ w \end{array} \right)$, with $u$ and $w$ denoting the horizontal and vertical velocity components respectively. In particular, we characterize the explicit form of the density $\rho$ and of the mass flux $\rho v$ inside the four layers in the following way. 

\smallskip
The deposit is characterized by a particles' volume fraction $\phi_{pack}$, therefore inside $\Omega_d$ we have that $\rho$ is given by the effective density $\rho_d = \phi_{pack} \rho_s + (1-\phi_{pack}) \rho_f$. Since we have assumed that the particles stop moving when they settle down, it is also natural to consider a zero mass flux $\rho v$ inside the deposit, i.e. we take $\rho v_d = \rho_d \left( \begin{array}{c} 0 \\ 0 \end{array} \right)$. Next, the solid volume fraction associated with the suspension is $\phi_s$, and thus the effective density inside $\Omega_m$ is given by $\rho_m = \phi_s \rho_s + (1-\phi_s) \rho_f$. Moreover, we assume that the sedimentation of particles onto the underlying deposit takes place in this layer solely in the vertical direction, oriented downward, with a velocity $-U_{sed}^\dyn$. Since the velocity field in non-zero here, we need to account for both the fluid's and the particles' velocities. Therefore, we may write the mass flux inside the suspension as $\rho v_m = \phi_s \rho_s \left( \begin{array}{c} u \ \ \ \ \\ w - U_{sed}^\dyn \end{array} \right) + (1-\phi_s) \rho_f \left( \begin{array}{c} u \\ w \end{array} \right)$. At last, there are no particles inside the fluid and the ambient air layers, and thus $\Omega_f$ and $\Omega_{\textrm{air}}$ can be simply characterized by the densities $\rho_f$ and $\rho_{\textrm{air}}$, respectively, and by the mass fluxes $\rho v_f = \rho_f \left( \begin{array}{c} U \\ W \end{array} \right)$ and $\rho v_{\textrm{air}} = \rho_{\textrm{air}} \left( \begin{array}{c} u \\ w \end{array} \right)$. Notice that $\rho_{\textrm{air}} \ll \rho_f$ and that, in order to avoid any confusion between the velocity of the fluid inside the domains $\Omega_f$ and $\Omega_m$, we have used the uppercase letters $U$ and $V$ to denote the water's velocity field in the upper pure fluid layer and the lowercase letters $u$ and $v$ to denote its counterpart in the suspension layer.

\smallskip
Introducing now the vector of moments $J = \left( \begin{array}{c} \rho \\ \rho v \end{array} \right)$, we can rewrite the continuity equation \eqref{eq:continuity-equation} in a more compact form as $\nabla_{(t,x,z)} \cdot J = 0$. Given any test function $\psi: \Omega \to \R \in C_c^1(\Omega)$, we then pass to a weak formulation of the latter to obtain
\begin{equation*}
\int_\Omega J \cdot \nabla_{(t,x,z)} \psi \dd \Omega = 0,
\end{equation*}
and an application of Stokes--Ostrogradsky theorem starting from the divergence of $\psi J$ over $\Omega$ leads to the relation
\begin{equation} \label{eq:main-relation}
\int_\Sigma \psi J \cdot \textbf{n} \dd \Sigma = 0,
\end{equation}
where we denote with a general $\Sigma$ the surface identified by the compact support of $\psi$ inside $\Omega$ and with $\mathbf{n}$ the outward normal vector to this boundary. The idea is to study each interface separately using these tools. For the interface $\Sigma_d^t$ between the deposit and the suspension, identified by the height $h_d(t,x)$, one may reduce the support of $\psi$ around $\Sigma_d^t$ so that we can rule out the influence of the other two interfaces and transform \eqref{eq:main-relation} into the equivalence
\begin{equation*}
\int_{\Sigma_d^t} \psi J_m \cdot \mathbf{n}_d \dd \Sigma_d^t - \int_{\Sigma_d^t} \psi J_d \cdot \mathbf{n}_d \dd \Sigma_d^t = 0,
\end{equation*}
where $\displaystyle \mathbf{n}_d = \frac{1}{\sqrt{1 + |\partial_x h_d|^2 + |\partial_t h_d|^2}} \left( \begin{array}{c} \partial_t h_d \\ \partial_x h_d \\ -1 \end{array} \right)$ is the outward normal vector to $\Sigma_d^t$, directed from $\Omega_d^t$ to $\Omega_m^t$. Choosing a suitable mollified step function $\psi$ such that $\psi \equiv 1$ on $\Sigma_d^t$, we can finally replace $J_m$ and $J_d$ with their explicit values to get
\begin{equation*}
(\rho_m - \rho_d) \partial_t h_d + \rho_m u |_{z = h_d} \partial_x h_d - \rho_m w |_{z = h_d} = - \phi_s \rho_s U_{sed}^\dyn.
\end{equation*}
But $\rho_m - \rho_d = (\phi_s - \phi_{pack}) (\rho_s - \rho_f)$ and simple algebraic manipulations lead to
\begin{equation*}
\partial_t h_d + \frac{\rho_m}{\rho_s - \rho_f} \frac{1}{\phi_s - \phi_{pack}} \left( u |_{z = h_d} \partial_x h_d - w |_{z = h_d} \right) = R_{sed}.
\end{equation*}
By assuming the non-penetration condition $u |_{z = h_d} \partial_x h_d = w |_{z = h_d}$ and recalling that $S_{sed}(t) = R_{sed} \chi(t > t_{ff})$, we finally recover equation \eqref{eq:interface 3} giving the evolution of the deposit's interface $\Sigma_d^t$ in terms of the height $h_d$. Obviously, the same reasoning can be applied to deal with the other two interfaces and recover equations \eqref{eq:interface 1}--\eqref{eq:interface 2}, by reducing relation \eqref{eq:main-relation} to $\Sigma_m^t$ and $\Sigma_f^t$ respectively and by using the definitions of $J_m$ and $J_f$, as done for $\Sigma_d^t$ and $J_m$, $J_d$. The only difference in these two cases is that we do not need to impose a non-penetration condition as for the interface $\Sigma_d^t$. Moreover, since $\rho_{\textrm{air}} \ll \rho_f$ it is natural to take $\rho_f - \rho_{\textrm{air}} \approx \rho_f$ when writing the spatio-temporal evolution of $\Sigma_f^t$ (this is the reason why we do not see the explicit appearance of the density $\rho_{\textrm{air}}$ in equation \eqref{eq:interface 1}, since the differences $\rho_f - \rho_{\textrm{air}}$ have been simply replaced by $\rho_f$).

\section{Derivation of the two-layer shallow water system} \label{app:model}

This second part is dedicated to the derivation of our model \eqref{eq:FL mass}--\eqref{eq:SL momentum}. Notice that we still focus on the case when $t > t_{ff}$, since we have assumed that no sedimentation takes place before this critical time and the shallow water system \eqref{eq:slumping-phase} accounting for the slumping phase has already been derived and extensively studied in previous works (see \cite{LarStaHin, DHLMS}). Starting from the assumptions presented in Section \ref{sec:model}, the main idea is to set up Euler- and Navier--Stokes-type equations to describe the evolution of the fluid and suspension layers respectively, and then derive in a suitable thin-layer approximation the expected shallow water system by depth-averaging the equations. Our derivation will follow similar steps and assumptions from the work of Gerbeau and Perthame \cite{GerPer}, and is also close in its reasoning to the earlier work of Liu and Mei \cite{LiuMei}. In what follows, we shall always use the convention that lowercase letters refer to quantities of the suspension layer, while capital letters refer to quantities of the fluid one. In particular, we introduce the horizontal $u = t(t,x,z)$ and vertical $w = w(t,x,z)$ components of the velocity field inside the suspension layer, together with their counterparts $U = U(t,x,z)$ and $W = W(t,x,z)$ inside the fluid layer. Moreover, recall that each of the two layers is contained between two interfaces, identified by lateral heights. In particular, the suspension layer spans from $h_d$ to $h_m$, while the fluid one spans from $h_m$ to $h_f$.

\smallskip
Then, the Euler system describing the evolution of the perfect fluid layer writes as
\begin{gather}
\partial_x U + \partial_z W = 0, \nonumber \\[4mm]
\rho_f (\partial_t U  + U \partial_x U + W \partial_z U) = - \partial_x P,
\label{eq:E fluid} \\[4mm]
\rho_f (\partial_t W + U \partial_x W + W \partial_z W) = - \partial_z P - \rho_f g, \nonumber
\end{gather}
while the suspension layer evolves following the Navier--Stokes system
\begin{gather}
\partial_x u + \partial_z w = 0, \nonumber \\[4mm]
\rho_m (\partial_t u + u \partial_x u + w \partial_z u) = - \partial_x p \hspace*{2cm} \label{eq:NS suspension} \\[2mm] 
\hspace*{2cm} + 2 \mu_m \partial_x^2 u + \mu_m \partial_{zx}^2 w + \mu_m \partial_z^2 u, \nonumber \\[4mm]
\rho_m (\partial_t w + u \partial_x w + w \partial_z w) = - \partial_z p - \rho_m g \hspace*{2cm} \nonumber \\[2mm]
\hspace*{2cm} + \mu_m \partial_{xz}^2 u + \mu_m \partial_x^2 w + 2 \mu_m \partial_z^2 w. \nonumber
\end{gather}
In the above equations, the quantities $P = P(t,x,z)$ and $p = p(t,x,z)$ denote the respective layers' pressures,
while $\mu_m$ defines the dynamic viscosity of the suspension. In particular, we may write $\mu_m = \rho_m \nu_m$, with $\nu_m$ defining the associated suspension's kinematic viscosity. This kind of approximation is reminiscent of the one proposed by Liu and Mei \cite{LiuMei}, but the distinction between layers is done here with the use of two different densities $\rho_f$ and $\rho_m$, rather than two different viscosities.

\smallskip
We complete the coupled systems \eqref{eq:E fluid}--\eqref{eq:NS suspension} with the equations governing the three interfaces \eqref{eq:interface 1}--\eqref{eq:interface 3} and with the following boundary conditions. At the interface with the deposit, we impose the Navier condition
\begin{equation} \label{eq:Navier condition}
\left( \kappa u - \mu_m \partial_z u \right) |_{z = h_d} = 0,
\end{equation}
and the non-penetration condition
\begin{equation} \label{eq:non-penetration at hd}
u|_{z = h_d} \partial_x h_d - w|_{z = h_d} = 0.
\end{equation}
Next, we assume continuity of the pressure and of the shear rate at the interface between suspension and fluid, namely
\begin{equation} \label{eq:continuity of pressure and shear}
p|_{z = h_m} = P|_{z = h_m}, \qquad \partial_z u|_{z = h_m} = \partial_z U|_{z = h_m},
\end{equation}
and we further impose the non-penetration condition
\begin{equation} \label{eq:non-penetration at hm}
u|_{z = h_m} \partial_x h_m - w|_{z = h_m} = U|_{z = h_m} \partial_x h_m - W|_{z = h_m}.
\end{equation}
Finally, at the upper interface between the fluid and the ambient air we impose the stress-free boundary condition
\begin{equation} \label{eq:stress-free condition}
(\sigma - P \mathbb{I}_2) \cdot \mathbf{n}_f = 0, \quad \textrm{at } z = h_f, 
\end{equation}
where $\sigma = \sigma(t,x,z)$ denotes the perfect fluid's deviatoric stress tensor
\begin{equation*}
\sigma = \left( \begin{array}{cc} 2 \partial_x U & \partial_x W + \partial_z U \\[2mm] \partial_x W + \partial_z U & 2 \partial_z W \end{array} \right),
\end{equation*}
the vector $\mathbf{n}_f$ denotes the outward normal at $z = h_f$, namely
\begin{equation*}
\mathbf{n}_f =  \frac{1}{\sqrt{1+|\partial_x h_f|^2}} \left( \begin{array}{c} -\partial_x h_f \\[2mm] 1 \end{array} \right),
\end{equation*}
and $\mathbb{I}_2$ is the identity matrix in two dimensions.

\smallskip
We proceed to determine the corresponding nondimensionalized form of the previous equations, by looking at a thin-layer regime of small ratio $\eps = \frac{h_0}{L_0} \ll 1$, where $h_0$ and $L_0$ denote respectively a characteristic height and length of the flow domain. Let us also introduce a characteristic speed $u_0$ of the flow and a unitary measure of dynamic viscosity $\mu$. Then, we consider the following rescaling of our relevant physical quantities
\begin{gather*}
x = L_0 x^*, \quad z = h_0 z^*, \quad t = \frac{L_0}{u_0} t^*, \\[2mm]
h_d = h_0 h_d^*, \quad h_m = h_0 h_m^*, \quad h_f = h_0 h_f^*, \\[4mm]
u = u_0 u^*, \quad U = u_0 U^*, \\[4mm]
w = \eps u_0 w^*, \quad W = \eps u_0 W^*, \quad U_{sed}^\dyn = \eps u_0 U_{sed}^*, \\[4mm]
p = \rho_m g h_0 p^*, \quad P = \rho_f g h_0 P^*, \\[4mm]
\sigma = \mu \frac{u_0}{h_0} \sigma^*, \quad \kappa = \rho_m \eps u_0 \kappa^*,
\end{gather*}
and we use the definitions $Re_m = \frac{h_0 u_0}{\nu_m}$ and $Fr = \frac{u_0}{\sqrt{g h_0}}$ for the Reynolds and Froude numbers of the flow. Notice that the term $\kappa u$ has the correct dimension of a friction and its rescaling is coherent with our choice of $\kappa$, since $\kappa/\rho_m \approx U_{sed}^\dyn = \eps u_0 U_{sed}^*$. The latter rescaling of the basal friction corresponds in particular to the one proposed in the work \cite{GerPer}.

\smallskip
By switching to the above nondimensional variables and by doing simple algebraic manipulations, the fluid layer's evolution system \eqref{eq:E fluid} rewrites
\begin{gather*}
\partial_x U + \partial_z W = 0, \nonumber \\[6mm]
\partial_t U  + U \partial_x U + W \partial_z U = - \frac{1}{Fr^2}\partial_x P, \nonumber \\[6mm]
\eps^2 (\partial_t W + U \partial_x W + W \partial_z W) = - \frac{1}{Fr^2} \partial_z P - \frac{1}{Fr^2}, \nonumber
\end{gather*}
where we have dropped the starred notation for convenience. Similarly, starting from \eqref{eq:NS suspension}, the nondimensionalized system governing the evolution of the suspension layer takes the form
\begin{gather*}
\partial_x u + \partial_z w = 0, \nonumber \\[2mm]
\partial_t u  + u \partial_x u + w \partial_z u = - \frac{1}{Fr^2}\partial_x p \hspace*{2cm} \nonumber \\[4mm]
\hspace*{2cm} + \frac{2 \eps}{Re_m} \partial_x^2 u + \frac{\eps}{Re_m} \partial_{zx}^2 w + \frac{1}{\eps Re_m} \partial_z^2 u, \nonumber \\[6mm]
\eps^2 (\partial_t w + u \partial_x w + w \partial_z w) = - \frac{1}{Fr^2} \partial_z p - \frac{1}{Fr^2} \hspace*{1cm} \nonumber \\[4mm]
\hspace*{3cm} + \frac{\eps}{Re_m} \partial_{xz}^2 u + \frac{\eps^3}{Re_m} \partial_x^2 w + \frac{2 \eps^2}{Re_m} \partial_z^2 w. \nonumber
\end{gather*}
Next, the structure of the equations for the three interfaces \eqref{eq:interface 1}--\eqref{eq:interface 3} does not modify when passing to nondimensional quantities. The Navier condition \eqref{eq:Navier condition} then rewrites as
\begin{equation} \label{eq:asymptotic Navier condition}
\eps \kappa u - \frac{1}{Re_m} \partial_z u  = 0, \quad \textrm{at } z = h_d,
\end{equation}
while the non-penetration condition \eqref{eq:non-penetration at hd} is left untouched by the rescaling. The same consideration obviously holds for the other non-penetration relation \eqref{eq:non-penetration at hm} at $z = h_m$. We continue with the continuity conditions \eqref{eq:continuity of pressure and shear} for pressure and shear rate at the suspension-fluid interface, which lead to the following nondimensional relations:
\begin{equation} \label{eq:continuity of pressure and shear}
\rho_m p|_{z = h_m} = \rho_f P|_{z = h_m}, \qquad \partial_z u|_{z = h_m} = \partial_z U|_{z = h_m}.
\end{equation}
At last, the rescaling of the stress-free condition \eqref{eq:stress-free condition} at the fluid-air interface leads to the relations
\begin{equation*}
\partial_z U = \left( 2 \eps^2 \partial_x U - \eps P \right) \partial_x h_f - \eps^2 \partial_x W, \quad \textrm{at } z = h_f,
\end{equation*}
and
\begin{equation*}
P = 2 \eps \partial_z W - \eps \left( \eps^2 \partial_x W + \partial_z U \right) \partial_x h_f, \quad \textrm{at } z = h_f.
\end{equation*}

Now, by collecting every term of lower order in $\eps$ in the previous rescaled equations and in the boundary conditions, we end up with an asymptotic regime where the fluid layer evolves through the system
\begin{gather}
\partial_x U + \partial_z W = 0, \label{eq:asymptotic E fluid 1} \\[6mm]
\partial_t U  + U \partial_x U + W \partial_z U = - \frac{1}{Fr^2}\partial_x P, \label{eq:asymptotic E fluid 2} \\[6mm]
\partial_z P + 1 = \mathcal{O}(\eps^2), \label{eq:asymptotic E fluid 3}
\end{gather}
while for the system governing the evolution of the suspension layer we recover the approximation
\begin{gather}
\partial_x u + \partial_z w = 0, \label{eq:asymptotic NS suspension 1} \\[6mm]
\partial_t u  + u \partial_x u + w \partial_z u = - \frac{1}{Fr^2}\partial_x p + \frac{1}{\eps Re_m} \partial_z^2 u + \mathcal{O}(\eps), \label{eq:asymptotic NS suspension 2} \\[6mm]
\partial_z p + 1 = \mathcal{O}(\eps). \label{eq:asymptotic NS suspension 3}
\end{gather}
The Navier condition along the top of the deposit carries over keeping the same structure and the same holds for the non-penetration condition \eqref{eq:non-penetration at hm} and the continuity of pressure and shear stress \eqref{eq:continuity of pressure and shear} at $z = h_m$.

\smallskip
The stress-free condition at the upper surface then provides the approximations
\begin{gather}
\partial_z U = \mathcal{O}(\eps^2), \quad \textrm{at } z = h_f, \label{eq:asymptotic continuity of shear} \\[4mm]
P = \mathcal{O}(\eps), \quad \textrm{at } z = h_f. \label{eq:asymptotic continuity of pressure}
\end{gather}
At last, recall that the equations for the evolving interfaces are not modified in the rescaling and thus correspond to the original ones \eqref{eq:interface 1}--\eqref{eq:interface 3}.

\smallskip
Starting from this asymptotic regime, the procedure for deriving the corresponding shallow water equations is standard. First of all, the hydrostatic relation \eqref{eq:asymptotic E fluid 3} tells us that the pressure $P$ does not depend on $x$ up to order $\mathcal{O}(\eps^2)$. Therefore, from the conservation equation for the vertical velocity \eqref{eq:asymptotic E fluid 2} and from the approximation \eqref{eq:asymptotic continuity of shear} at $z = h_f$ we deduce that $\partial_z U = \mathcal{O}(\eps^2)$ inside the whole fluid layer (and in particular at its lower interface $z = h_m$). Next, multiplying by $\eps$ both sides of equation \eqref{eq:asymptotic NS suspension 2}, we find that $\partial_z^2 u = \mathcal{O}(\eps)$. Moreover, the Navier condition \eqref{eq:asymptotic Navier condition} implies that $\partial_z u|_{z = h_d} = \mathcal{O}(\eps)$, while the transmission condition for the shear stress \eqref{eq:continuity of pressure and shear} across $z = h_m$ gives $\partial_z u|_{z = h_m} = \mathcal{O}(\eps^2)$. All these considerations allow to finally infer that the flow behaves like a plug in both the fluid and the suspension layers, and does not depend on $z$ at the leading order in $\eps^2$ and $\eps$ respectively. In particular, we deduce the existence of two velocities $\overline{U} = \overline{U}(t,x)$ and $\overline{u} = \overline{u}(t,x)$ such that
\begin{equation} \label{eq:plug flow}
U(t,x,z) = \overline{U}(t,x) + \mathcal{O}(\eps^2), \quad u(t,x,z) = \overline{u}(t,x) + \mathcal{O}(\eps).
\end{equation}

We now proceed by integrating in $z$ over $[h_m, h_f]$ the fluid's incompressibility condition \eqref{eq:asymptotic E fluid 1} and we apply Leibniz integral rule to recover
\begin{equation*}
\partial_x \int_{h_m}^{h_f} U(t,x,z) \dd z - U|_{z = h_f} \partial_x h_f + W|_{z = h_f} \\[2mm] + U|_{z = h_m} \partial_x h_m - W|_{z = h_m} = 0.
\end{equation*}
Using the equations \eqref{eq:interface 1}--\eqref{eq:interface 2} for the interfaces $\Sigma_f^t$ and $\Sigma_m^t$, together with the non-penetration condition \eqref{eq:non-penetration at hm}, we thus recover the mass conservation equation for the fluid layer
\begin{equation*}
\partial_t H + \partial_x (H \bar{U}) = \left(\frac{\phi_{pack}}{\phi_s} - 1 \right) R_{sed},
\end{equation*}
where we have defined the height of the fluid layer $H = h_f - h_m$, which simply implies that $\int_{h_m}^{h_f} U(t,x,z) \dd z = H \bar{U}$ at the leading order. Taking the limit $\eps \to 0$ in the hydrostatic relation \eqref{eq:asymptotic E fluid 3} and integrating over $[z,h_f]$, we then use the boundary condition \eqref{eq:asymptotic continuity of pressure} for $P$ at $z = h_f$ to infer the explicit form of the pressure in the fluid layer, which reads
\begin{align*}
P(t,x,z) & = h_f - z \\[2mm]
& = H + h + h_d - z, \quad z \in [h_m, h_f].
\end{align*}
where we have further defined the height of the suspension layer $h = h_m - h_d$. In particular, we find that $\partial_x P = \partial_x h_f = \partial_x (H+h+h_d)$. Therefore, one proceed by integrating in $z$ over $[h_m,h_f]$ the equation for the conservation of horizontal velocity \eqref{eq:asymptotic E fluid 2} to initially obtain
\begin{equation*}
\int_{h_m}^{h_f} \partial_t U \dd z + \int_{h_m}^{h_f} 2 U \partial_x U \dd z
+ U|_{z = h_f} W|_{z = h_f} - U|_{z = h_m} W|_{z = h_m}
= -\frac{1}{Fr^2} H \partial_x (H + h + h_d),
\end{equation*}
where we have integrated by parts the nonlinear term $W \partial_z U$ and used the incompressibility condition \eqref{eq:asymptotic E fluid 1} to see that $\partial_z W = - \partial_x U$. By now applying Leibniz integral rule to both the temporal derivative $\partial_t U$ and the nonlinear term $2 U \partial_x U$, we then recover
\begin{multline*}
\partial_t \int_{h_m}^{h_f} U(t,x,z) \dd z + \partial_x \int_{h_m}^{h_f} U^2(t,x,z) \dd z
- U|_{z = h_f} \left(\partial_t h_f + U|_{z = h_f} \partial_x h_f - W|_{z = h_f} \right) \\[4mm]
+  U|_{z = h_m} \left(\partial_t h_m + U|_{z = h_m} \partial_x h_m - W|_{z = h_m} \right)
= -\frac{1}{Fr^2} H \partial_x (H + h + h_d),
\end{multline*}
which finally provides the momentum conservation equation for the fluid layer (going back to dimensional variables)
\begin{equation*}
 \partial_t (H \bar{U}) + \partial_x (H \bar{U}^2) = \left(\frac{\phi_{pack}}{\phi_s} - 1 \right) R_{sed} \bar{U}
 - g H \partial_x (H + h + h_d),
\end{equation*} 
by replacing the expressions for the interfaces' equations \eqref{eq:interface 1}--\eqref{eq:interface 2}, combined with the non-penetration condition \eqref{eq:non-penetration at hm} at $z = h_m$, and by using the expansion \eqref{eq:plug flow} for $U$ which tells us that
\begin{equation*}
\int_{h_m}^{h_f} U^2(t,x,z) \dd z = H \bar{U}^2, \quad U|_{z = h_f} = U|_{z = h_m} = \bar{U},
\end{equation*}

\smallskip
We continue with the analysis of the asymptotic system governing the suspension layer. We start by integrating in $z$ over $[h_d,h_m]$ the incompressibility relation \eqref{eq:asymptotic NS suspension 1} and using Leibniz rule as before to get
\begin{equation*}
\partial_x \int_{h_d}^{h_m} u(t,x,z) \dd z - u|_{z = h_m} \partial_x h_m + w|_{z = h_m} \\[2mm] + u|_{z = h_d} \partial_x h_d - w|_{z = h_d} = 0,
\end{equation*}
from which we easily derive the mass conservation equation for the suspension layer
\begin{equation*}
\partial_t h + \partial_x (h \bar{u}) = - \frac{\phi_{pack}}{\phi_s} R_{sed},
\end{equation*}
by using the equations \eqref{eq:interface 2}--\eqref{eq:interface 3} for the interfaces $\Sigma_m^t$ and $\Sigma_d^t$, and by recalling the definition $h = h_m - h_d$, so that the expansion \eqref{eq:plug flow} leads to $\int_{h_d}^{h_m} u(t,x,z) \dd z = h \bar{u}$. We then consider the hydrostatic relation \eqref{eq:asymptotic NS suspension 3} and integrate it over $[z,h_m]$ to exploit the transmission condition for the pressure \eqref{eq:continuity of pressure and shear} across the interface $\Sigma_m^t$ to find an explicit expression for $p$, namely
\begin{align*}
p(t,x,z) & = \frac{\rho_f (h_f - h_m) + \rho_m (h_m - z)}{\rho_m} \\[4mm]
& = \frac{\rho_f}{\rho_m} H + h + h_d - z, \quad z \in [h_m, h_f].
\end{align*}
Once again, from the above we infer that $\partial_x p = \partial_x \left( \frac{\rho_f}{\rho_m} H + h + h_d \right)$, and this relation can be injected in the conservation equation for the suspension's horizontal velocity \eqref{eq:asymptotic NS suspension 2} to recover, after integration in $z$ over $[h_d,h_m]$, the initial approximation
\begin{multline*}
\int_{h_d}^{h_m} \partial_t u \dd z + \int_{h_d}^{h_m} u \partial_x u \dd z + \int_{h_d}^{h_m} w \partial_z u \dd z \\[4mm]
= -\frac{1}{Fr^2} h \partial_x \left( \frac{\rho_f}{\rho_m} H + h + h_d \right) + \frac{1}{\eps Re_m} \int_{h_d}^{h_m} \partial_z^2 u \dd z + \mathcal{O}(\eps).
\end{multline*}
Now, from the Navier condition \eqref{eq:asymptotic Navier condition} and the continuity of the shear rate \eqref{eq:continuity of pressure and shear} at $z = h_m$, we infer that
\begin{equation*}
\frac{1}{\eps Re_m} \int_{h_d}^{h_m} \partial_z^2 u \dd z = - \kappa u|_{z = h_d} + \mathcal{O}(\eps),
\end{equation*}
and we proceed with the same calculations used previously for the fluid layer, to end up with the following equation:
\begin{multline*}
\partial_t \int_{h_d}^{h_m} u(t,x,z) \dd z + \partial_x \int_{h_d}^{h_m} u^2(t,x,z) \dd z
- u|_{z = h_m} \left(\partial_t h_m + u|_{z = h_m} \partial_x h_m - w|_{z = h_m} \right) \\[4mm]
+ u|_{z = h_d} \left(\partial_t h_d + u|_{z = h_d} \partial_x h_d - w|_{z = h_d} \right)
= -\frac{1}{Fr^2} h \partial_x \left( \frac{\rho_f}{\rho_m} H + h + h_d \right) - \kappa u|_{z = h_d} + \mathcal{O}(\eps).
\end{multline*}
From this approximation, one finally deduces the momentum conservation equation (in dimensional form) for the suspension layer
\begin{equation*}
\partial_t (h \bar{u}) + \partial_x (h \bar{u}^2) = - \frac{\phi_{pack}}{\phi_s} R_{sed} \bar{u} - \kappa \bar{u} \\[2mm]
- g h \partial_x \left( \frac{\rho_f}{\rho_m} H + h + h_d \right),
\end{equation*}
where we have replaced the expressions for the interfaces' equations \eqref{eq:interface 2}--\eqref{eq:interface 3}, together with the non-penetration condition \eqref{eq:non-penetration at hd} at $z = h_d$ to get rid of the term $u|_{z = h_d} \partial_x h_d - w|_{z = h_d}$, and used once again the expansion \eqref{eq:plug flow} for $u$ to see that
\begin{equation*}
\int_{h_d}^{h_m} u^2(t,x,z) \dd z = h \bar{u}^2, \quad u|_{z = h_m} = u|_{z = h_d} = \bar{u}.
\end{equation*}


\nocite{*}
\bibliographystyle{plain}
\bibliography{Bibliography_Three-Layer-Model}

\setlength\parindent{0pt}

\end{document}